\documentclass[12pt]{article}
\usepackage{graphicx}
\usepackage{longtable}
\usepackage{enumerate}
\usepackage{multirow}%
\usepackage{amsmath,amssymb,amsfonts}%
\usepackage{amsthm}%
\usepackage{mathrsfs}%
\usepackage[title]{appendix}%
\usepackage{xcolor}%
\usepackage{subfigmat, subfigure,etoolbox,float}%
\usepackage{textcomp}%
\usepackage{manyfoot}%
\usepackage{booktabs}%
\usepackage{algorithm}%
\usepackage{algorithmicx}%
\usepackage{algpseudocode}%
\usepackage{listings}%
\usepackage[round]{natbib}
\usepackage{url} 

\DeclareMathAlphabet      {\mathbfit}{OML}{cmm}{b}{it}

\theoremstyle{thmstyleone}%
\newtheorem{theorem}{Theorem}
\newtheorem{proposition}[theorem]{Proposition}%

\theoremstyle{thmstyletwo}%

\theoremstyle{thmstylethree}%

\begin{document}

\def\spacingset#1{\renewcommand{\baselinestretch}%
{#1}\small\normalsize} \spacingset{1}

\allowdisplaybreaks

\title{Cluster weighted models with multivariate skewed distributions for functional data
\thanks{This work was supported by the Natural Sciences and Engineering Research Council of Canada under Grant DG-2018-04449.}
}


\author{Cristina Anton,$^1$  Roy Shivam Ram Shreshtth$^2$} 
\date{$^1$Department of Mathematics and Statistics, MacEwan University, 
103C, 10700-104 Ave., Edmonton, AB T5J 4S2, Canada, email: \texttt{popescuc@macewan.ca}\\
$^2$Department of Mathematics and Statistics, Indian Institute of Technology Kanpur}
\maketitle
\begin{abstract}
We propose a  clustering method, funWeightClustSkew, based on mixtures of  functional linear regression models and three skewed multivariate distributions: the variance-gamma distribution, the  skew-t distribution, and the normal-inverse Gaussian distribution. Our approach follows the framework of the functional high dimensional data clustering (funHDDC) method, and we extend to functional data the cluster weighted models based on skewed distributions used for finite dimensional multivariate data.  We consider several parsimonious models, and to estimate the parameters we construct an expectation maximization (EM) algorithm. We illustrate the performance of funWeightClustSkew for  simulated data and for the Air Quality dataset. 
\end{abstract}

\noindent%


\section{Introduction}\label{sec1} 
Smart devices and other modern technologies record huge amounts of data measured continuously in time.  These data are better represented as curves instead of  finite-dimensional vectors, and they are analyzed using statistical methods specific to functional data \citep{RamsaySilverman:2006, FerratyVieu:2006,HorvathKokoszka:2012}.  Many times  more than one curve is collected for one individual, e.g. some smart watches can measure the heart rate, blood oxygen saturation, skin temperature, calories burned and other activity metrics. Determining  homogeneous groups of data is the first step for complex applications that  require the analysis of a huge amount of functional data. When a linear regression relationship exists between variables, it is useful to include it in the clustering model. 

Here we propose a  clustering method, funWeightClustSkew, for multivariate functional data based on functional linear regression mixture models and skewed multivariate distributions. Our approach uses the framework of the funHDDC method \citep{SchmutzJacquesBouveyronChezeMartin:2020} to extend the multivariate cluster weighted models (CWM) with skewed distributions \citep{GallaugherTomarchioMcNicholasPunzo:2022} to functional data. We consider functional linear regression models specific to each cluster, with multivariate functional response and predictors. We use multivariate functional principal component analysis (MFPCA) and we assume that the scores have the variance-gamma (VG) distribution, the  skew-t (ST) distribution, or the normal-inverse Gaussian (NIG) distribution. We consider several parsimonious models and we construct a version of the EM algorithm to estimate the parameters. This paper is also an extension of the work in \cite{AntonSmith:2025} where we use a similar approach but we assume multivariate normal distribution instead of the skewed VG, ST, or NIG distributions. 

CWMs, first introduced in \cite{Gershenfeld:1997}, add flexibility by  taking into account the distribution of the covariates. In addition to both response and covariates with normal distributions \citep{DangPunzoMcNicholasIngrassiaBrowne:2017},  some extensions consider  t-distributions \citep{IngrassiaMinottiVittadini:2012}, contaminated normal distributions \citep{PunzoMcNicholas:2017},  and various types of response variables and  covariates of mixed-type (\citealp{IngrassiaPunzoVittadiniMinotti:2015},  \citealp{PunzoIngrassia:2016}). The most common distributions of the exponential family and the t-distribution are included in the R package {\it flexCWM} \citep{MazzaPunzoIngrassia:2018}. A few CWMs consider multivariate responses \citep{DangPunzoMcNicholasIngrassiaBrowne:2017, PunzoMcNicholas:2017,  GallaugherTomarchioMcNicholasPunzo:2022} and recently matrix-variate CWMs are constructed in \cite{TomarchioMcNicholasPunzo:2021}.
 
To the best of our knowledge there are not many papers that consider mixtures of functional linear regression models for clustering. In \cite{Chiou:2012} a classification method  based on subspace projection is proposed, but only for linear functional regression models with a single functional response and predictor. Functional principal components analysis is used in \cite{YaoFuLee:2011} for a clustering method based on a  functional regression model with a scalar response and only one functional predictor. In \cite{CondeTavakoliEzer:2021}  a cluster-specific model with a single functional response and multiple functional predictors is used for finding clusters in a gene expression time course data set. A mixture constructed with concurrent  functional linear models, is used in \cite{WangHuangWuYao:2016} to analyze the CO$_2$ emissions - GDP relationship. In \cite{Chamroukhi:2016}  clustering is done using models based on polynomial,  spline,  or B-spline regression mixtures with a single functional response and multiple functional predictors.

The paper is organized as follows. Results about the multivariate skewed distributions are presented in Section \ref{sectskew}. In Section \ref{section2} we construct the cluster
weighted models for functional data.  Section \ref{section3} includes the EM algorithm to estimate the parameters. Applications for both simulated
and real-world data are presented in Section \ref{section4}. In Section \ref{section5} we provide concluding remarks and future directions.


   \section{Multivariate skewed distributions}
 \label{sectskew}
 We present some preliminary results regarding the distributions considered in this paper. Following \cite{GallaugherTomarchioMcNicholasPunzo:2022} we start with the generalized inverse Gaussian (GIG) distribution, and then we summarize the properties of three multivariate skewed distributions: the  VG, the  ST,  and the NIG distributions.  
 
 The probability density function (pdf) of the $GIG(a, b, \lambda)$ distribution \citep{Jorgensen:2012} with parameters $a>0$, $b>0$, and $\lambda\in \mathbb{R}$ is 
 \begin{equation*}
 h(w;a,b,\lambda)=\left(\frac{a}{b}\right)^{\frac{\lambda}{2}}\frac{w^{\lambda-1}}{2K_\lambda(\sqrt{ab}}\exp\left(-\frac{1}{2}\left(aw+\frac{b}{w}\right)\right),
 \end{equation*} 
 where 
 \begin{equation*}
 K_\lambda(u)=\frac{1}{2}\int_0^\infty w^{\lambda-1}\exp\left(-\frac{u}{2}\left(w+\frac{1}{w}\right)\right)dw
 \end{equation*}
 is the modified Bessel function of the third kind with index $\lambda$. The following formulas \citep{BrowneMcNicholas:2015} for the expectations of some functions of $W\sim GIG(a, b, \lambda)$ are used in the next sections for parameter estimation:
 \begin{align}
& E[W]=\sqrt{\frac{b}{a}}\frac{K_{\lambda+1}(\sqrt{ab})}{K_{\lambda}(\sqrt{ab})},\label{es1}\\
&E\left[\frac{1}{W}\right]=\sqrt{\frac{a}{b}}\frac{K_{\lambda+1}(\sqrt{ab})}{K_{\lambda}(\sqrt{ab})}-\frac{2\lambda}{b},\label{es2}\\
&E[\log(W)]=\log\left(\sqrt{\frac{b}{a}}\right)+\frac{1}{K_{\lambda}(\sqrt{ab})}\frac{\partial}{\partial \lambda}K_{\lambda}(\sqrt{ab}).\label{es3}
 \end{align}
Setting  $\omega=\sqrt{ab}>0$, $\eta=\sqrt{b/a}>0$, in \cite{BrowneMcNicholas:2015}  the following alternative parametrization is proposed
\begin{equation}
h(w;\omega,\eta,\lambda)=\left(\frac{\omega}{\eta}\right)^{\lambda-1}\frac{1}{2\eta K_\lambda(\omega)}\exp\left(-\frac{\omega}{2}\left(\frac{\omega}{\eta}+\frac{\eta}{\omega}\right)\right).
\end{equation} 
We denote the $GIG$ distribution with this parametrization by $I(\omega, \eta,\lambda)$. 

Many skewed distributions can be obtained using the following normal variance-mean mixture model \citep{GallaugherTomarchioMcNicholasPunzo:2022}:
\begin{equation}
\mathbfit{V}=\boldsymbol{\mu}+W\boldsymbol{\alpha}+\sqrt{W}\mathbfit{U}.\label{skewm}
\end{equation} 
Here $\mathbfit{V}$ is a $d-$variate random variable with a skewed distribution, $\boldsymbol{\mu}$ is the location parameter, $\boldsymbol{\alpha}$ is the skewness parameter,  
 $W$ is a positive random variable, and $\mathbfit{U}\sim N({\bf 0}, \boldsymbol{\Sigma})$, where $N({\bf 0}, \boldsymbol{\Sigma})$ denotes a $d-$variate normal distribution.

As in  \cite{GallaugherTomarchioMcNicholasPunzo:2022} we consider the VG, the ST and the NIG distributions.
We denote
\begin{equation*}
\delta(\mathbfit{v};\boldsymbol{\mu}, \boldsymbol{\Sigma}):=\left(\mathbfit{v}-\boldsymbol{\mu}\right)^\top\boldsymbol{\Sigma}^{-1}\left(\mathbfit{v}-\boldsymbol{\mu}\right),~\rho(\boldsymbol{\alpha,\Sigma}):=\boldsymbol{\alpha^\top\Sigma^{-1}\alpha}.
\end{equation*} 
\begin{itemize}
\item The $d-$ variate variance-gamma distribution $VG_d(\boldsymbol{\mu},\boldsymbol{\alpha},\boldsymbol{\Sigma},\psi)$, $\psi>0$, can be obtained from \eqref{skewm} when $W\sim G(\psi,\psi)$, where $G(a,b)$, $a>0$, $b>0$ is the Gamma distribution with pdf
\begin{equation}
h(w;a,b)=\frac{b^a}{\Gamma(a)}w^{a-1}\exp\left(-bw\right).
\label{h1l}
\end{equation}
For the VG distribution we have
\begin{equation*}
W| \mathbfit{V}=\mathbfit{v}\sim GIG\left(\rho(\boldsymbol{\alpha, \Sigma})+2\psi, \delta(\mathbfit{v};\boldsymbol{\mu}, \boldsymbol{\Sigma}), \psi-d/2\right).
\end{equation*}
The pdf of the $VG_d(\boldsymbol{\mu},\boldsymbol{\alpha},\boldsymbol{\Sigma},\psi)$ distribution is
\begin{align}
f_{VG}(\mathbfit{v};\boldsymbol{\mu,\alpha,\Sigma},\psi)&=\frac{2\psi^\psi \exp\left(\left(\mathbfit{v}-\boldsymbol{\mu}\right)^\top\boldsymbol{\Sigma}^{-1}\boldsymbol{\alpha}\right)}{\left(2\pi\right)^{d/2}|\boldsymbol{\Sigma}|^{1/2}\Gamma(\psi)}\left(\frac{\delta(\mathbfit{v};\boldsymbol{\mu}, \boldsymbol{\Sigma})}{\rho(\boldsymbol{\alpha,\Sigma})+2\psi}\right)^{\frac{\psi-d/2}{2}}\notag\\
&\times K_{\psi-d/2}\left(\sqrt{\left(\rho(\boldsymbol{\alpha,\Sigma})+2\psi\right)\delta(\mathbfit{v};\boldsymbol{\mu, \Sigma})}\right).\label{pdfVG}
\end{align}
\item The $d-$variate skew-t distribution $ST_d(\boldsymbol{\mu},\boldsymbol{\alpha},\boldsymbol{\Sigma},\nu)$, $\nu>0$, can be obtained from \eqref{skewm} when $W\sim IG(\nu/2,\nu/2)$, where $IG(a,b)$, $a>0$, $b>0$ is the inverse Gamma distribution with pdf
\begin{equation}
h(w;a,b)=\frac{b^a}{\Gamma(a)}w^{-a-1}\exp\left(-\frac{b}{w}\right).
\label{h2l}
\end{equation}
For the ST distribution we have
\begin{equation*}
W| \mathbfit{V}=\mathbfit{v}\sim GIG\left(\rho(\boldsymbol{\alpha, \Sigma}), \delta(\mathbfit{v};\boldsymbol{\mu}, \boldsymbol{\Sigma})+\nu, -(\nu+d)/2\right).
\end{equation*}
The pdf of the $ST_d(\boldsymbol{\mu},\boldsymbol{\alpha},\boldsymbol{\Sigma},\nu)$ distribution is
\begin{align}
f_{ST}(\mathbfit{v};\boldsymbol{\mu,\alpha,\Sigma},\nu)&=\frac{2\left(\frac{\nu}{2}\right)^{\nu /2}\exp\left(\left(\mathbfit{v}-\boldsymbol{\mu}\right)^\top\boldsymbol{\Sigma}^{-1}\boldsymbol{\alpha}\right)}{\left(2\pi\right)^{d/2}|\boldsymbol{\Sigma}|^{1/2}\Gamma\left(\frac{\nu}{2}\right)}\left(\frac{\delta(\mathbfit{v};\boldsymbol{\mu}, \boldsymbol{\Sigma})+\nu}{\rho(\boldsymbol{\alpha,\Sigma})}\right)^{-\frac{\nu+d}{4}}\notag\\
&\times K_{-\frac{\nu+d}{2}}\left(\sqrt{\rho(\boldsymbol{\alpha,\Sigma})\left(\delta(\mathbfit{v};\boldsymbol{\mu, \Sigma})+\nu\right)}\right).\label{pdfST}
\end{align}
\item The $d-$variate normal-inverse Gaussian distribution $NIG_d(\boldsymbol{\mu},\boldsymbol{\alpha},\boldsymbol{\Sigma},\kappa)$, $\kappa>0$, can be obtained from \eqref{skewm} when $W\sim IN(1,\kappa)$, where $IN(a,b)$, $a>0$, $b>0$ is the inverse Gaussian distribution with pdf
\begin{equation}
h(w;a,b)=\frac{a}{\sqrt{2\pi}}\exp(ab)w^{-3/2}\exp\left(-\frac{1}{2}\left(\frac{a^2}{w}+b^2w\right)\right).
\label{h3l}
\end{equation}
Notice that $W\sim IN(1,\kappa)$ is a special case of $W\sim I(\omega,1,\lambda)$, where $\omega=\kappa$, $\lambda=-1/2$.
For the NIG distribution we have
\begin{equation*}
W| \mathbfit{V}=\mathbfit{v}\sim GIG\left(\rho(\boldsymbol{\alpha, \Sigma})+\kappa^2, \delta(\mathbfit{v};\boldsymbol{\mu}, \boldsymbol{\Sigma})+1, -(1+d)/2\right).
\end{equation*}
The pdf of the $NIG_d(\boldsymbol{\mu},\boldsymbol{\alpha},\boldsymbol{\Sigma},\kappa)$ distribution is
\begin{align}
f_{NIG}(\mathbfit{v};\boldsymbol{\mu,\alpha,\Sigma},\kappa)&=\frac{2\exp\left(\left(\mathbfit{v}-\boldsymbol{\mu}\right)^\top\boldsymbol{\Sigma}^{-1}\boldsymbol{\alpha}+\kappa\right)}{\left(2\pi\right)^{(d+1)/2}|\boldsymbol{\Sigma}|^{1/2}}\left(\frac{\delta(\mathbfit{v};\boldsymbol{\mu}, \boldsymbol{\Sigma})+1}{\rho(\boldsymbol{\alpha,\Sigma})+\kappa^2}\right)^{-\frac{1+d}{4}}\notag\\
&\times K_{-\frac{1+d}{2}}\left(\sqrt{\left(\rho(\boldsymbol{\alpha,\Sigma})+\kappa^2\right)\left(\delta(\mathbfit{v};\boldsymbol{\mu, \Sigma})+1\right)}\right).\label{pdfNIG}
\end{align}
\end{itemize}
\section{Multivariate functional cluster weighted model }
\label{section2}
For any compact interval  $\mathcal{T}$ in $\mathbb{R}$, we consider the Hilbert space $L^2(\mathcal{T})$ $=\{f:\mathcal{T}\rightarrow \mathbb{R}, \int_{\mathcal{T}} f^2(t)dt<\infty\}$ with the inner product $<f,g>=\int_{\mathcal{T}} f(t)g(t)dt$ and the norm $\|f\|=<f,f>^{1/2}$ \citep{RamsaySilverman:2006}.

We assume that the  $n$ $p_Y$-variate response curves $\{{\mathbfit{Y}}_1,\ldots,{\mathbfit{Y}}_n\}$ are independent realizations of a $L^2$- continuous stochastic process $\mathbfit{Y}=\{{\mathbfit{Y}}(t)\}_{t\in \mathcal{T}_Y}$ $=\{(Y^1(t),\ldots,Y^{p_Y}(t))^\top\}_{t\in\mathcal{T}_Y}\in \mathbb{H}_Y$, where $ \mathcal{T}_Y\subset \mathbb{R}$ is a compact interval  and $\mathbb{H}_Y:=\{\mathbfit{f}=(f_1, \ldots,f_{p_Y})^\top:\mathcal{T}_Y\rightarrow \mathbb{R}^{p_Y}, f_i\in L^2(\mathcal{T}_Y), i=1, \ldots, p_Y\}$ is a Hilbert space with the inner product $<\mathbfit{f},\mathbfit{g}>_{\mathbb{H}_Y}=\sum_{l=1}^{p_Y}<f_l,g_l>$ and the norm $\|\mathbfit{f}\|_{\mathbb{H}_Y}=<\mathbfit{f},\mathbfit{f}>_{\mathbb{H}_Y}^{1/2}$.

Similarly we assume that the  $n$ $p_X$-variate covariate curves $\{{\mathbfit{X}}_1,\ldots,{\mathbfit{X}}_n\}$ are independent realizations of a $L^2$- continuous stochastic process $\mathbfit{X}=\{{\mathbfit{X}}(t)\}_{t\in \mathcal{T}_X}$ $=\{(X^1(t),\ldots,$ $X^{p_X}(t))^\top\}_{t\in \mathcal{T}_X}$ $\in \mathbb{H}_X$, where $ \mathcal{T}_X\subset \mathbb{R}$ is a compact interval  and $\mathbb{H}_X:=\{\mathbfit{f}=(f_1, \ldots,f_{p_X})^\top:\mathcal{T}_X\rightarrow \mathbb{R}^{p_X}, f_i\in L^2(\mathcal{T}_X), i=1, \ldots, p_X\}$ is a Hilbert space with the inner product $<\mathbfit{f},\mathbfit{g}>_{\mathbb{H}_X}=\sum_{j=1}^{p_X}<f_j,g_j>$ and the norme $\|\mathbfit{f}\|_{\mathbb{H}_X}=<\mathbfit{f},\mathbfit{f}>_{\mathbb{H}_X}^{1/2}$.

For each pair of curves $({\mathbfit{Y}}_i, {\mathbfit{X}}_i)$ we have access to a finite set of values $y_{i}^{s_Y}(t_{i1}^Y)\ldots,y_{i}^{s_Y}(t_{im_i}^Y)$, $x_{i}^{s_X}(t_{i1}^X)\ldots,x_{i}^{s_X}(t_{in_i}^X)$, where $t_{i1}^Y<t_{i2}^Y<\cdots<t_{im_i}^Y$, $t_{i1}^X<t_{i2}^X<\cdots<t_{in_i}^X$, $t_{ij}^Y\in \mathcal{T}_Y$, $t_{il}^X\in \mathcal{T}_X$, $j=1,\ldots, m_i$, $l=1,\ldots, n_i$, $s_Y=1,\ldots, p_Y$, $s_X=1,\ldots, p_X$, $i=1,\ldots, n$. To reconstruct the functional form of the data we assume that the curves belong to a finite dimensional space, and we have:
\begin{align}
Y_i^l(t)&=\sum_{r=1}^{R_l^Y} c_{Y,ir}^l\xi_{Y,r}^l(t), \quad X_i^j(t)=\sum_{r=1}^{R_j^X} c_{X,ir}^j\xi_{X,r}^j(t).\label{eqfirst}
\end{align} 
Here $\{\xi_{Y,r}^l\}_{1\le r\le R_l^Y}$ is the basis for the $l^{th}$ components of the multivariate curves $\{{\mathbfit{Y}}_1,\ldots,{\mathbfit{Y}}_n\}$, $c_{Y,ir}^l$ are the coefficients, and $R_l^Y$ is the number of basis functions. Similarly for the covariate curves $\{{\mathbfit{X}}_1,\ldots,{\mathbfit{X}}_n\}$,  $\{\xi_{X,r}^j\}_{1\le r\le R_j^X}$ is the basis for the $j^{th}$ components, $c_{X,ir}^j$ are the coefficients, and $R_j^X$ is the number of basis functions.

Following \cite{SchmutzJacquesBouveyronChezeMartin:2020}, we rewrite \eqref{eqfirst} as
\begin{align}
{\mathbfit{Y}}(t)&={\mathbfit{C}_Y}\boldsymbol{\xi}_Y^\top (t),\quad {\mathbfit{Y}}(t)=({\mathbfit{Y}}_1(t),\ldots,{\mathbfit{Y}}_n(t))^\top, \label{eqsecond}\\
{\mathbfit{X}}(t)&={\mathbfit{C}_X}\boldsymbol{\xi}_X^\top (t),\quad {\mathbfit{X}}(t)=({\mathbfit{X}}_1(t),\ldots,{\mathbfit{X}}_n(t))^\top, \label{eqsecondbis}
\end{align}
 with
 \begin{align*}
 \mathbfit{C_Y}&=\begin{pmatrix}
 c_{Y,11}^1&\ldots&c_{Y,1R_1^Y}^1&c_{Y,11}^2&\ldots&c_{Y,1R_2^Y}^2&\ldots& c_{Y,11}^{p_Y}&\ldots&c_{Y,1R_p^Y}^{p_Y}\\
 \vdots&\ddots&\vdots&\vdots&\ddots&\vdots&\ldots&\vdots&\ddots&\vdots\\
 c_{Y,n1}^1&\ldots&c_{Y,nR_1^Y}^1&c_{Y,n1}^2&\ldots&c_{Y,nR_2^Y}^2&\ldots &c_{Y,n1}^{p_Y}&\ldots&c_{Y,nR_p^Y}^{p_Y}
 \end{pmatrix},\\
 \boldsymbol{\xi}_Y(t)&=\begin{pmatrix}
 \xi_{Y,1}^1(t)&\ldots&\xi_{Y,R_1^Y}^1&0&\ldots&0&\ldots& 0&\ldots&0\\
 0&\ldots&0&\xi_{Y,1}^2(t)&\ldots&\xi_{Y,R_2^Y}^2(t)&\ldots& 0&\ldots&0\\
 \vdots&\ddots&\vdots&\vdots&\ddots&\vdots&\ldots&\vdots&\ddots&\vdots\\
 0&\ldots&0&0&\ldots&0&\ldots &\xi_{Y,1}^{p_Y}(t)&\ldots&\xi_{Y,R_{p_Y}^Y}^{p_Y}(t)
 \end{pmatrix},\\
 \mathbfit{C_X}&=\begin{pmatrix}
 c_{X,11}^1&\ldots&c_{X,1R_1^X}^1&c_{X,11}^2&\ldots&c_{X,1R_2^X}^2&\ldots& c_{X,11}^p&\ldots&c_{X,1R_p^X}^{p_X}\\
 \vdots&\ddots&\vdots&\vdots&\ddots&\vdots&\ldots&\vdots&\ddots&\vdots\\
 c_{X,n1}^1&\ldots&c_{X,nR_1^X}^1&c_{X,n1}^2&\ldots&c_{X,nR_2^X}^2&\ldots &c_{X,n1}^p&\ldots&c_{X,nR_p^X}^{p_X}
 \end{pmatrix},\\
 \boldsymbol{\xi}_X(t)&=\begin{pmatrix}
 \xi_{X,1}^1(t)&\ldots&\xi_{X,R_1^X}^1&0&\ldots&0&\ldots& 0&\ldots&0\\
 0&\ldots&0&\xi_{X,1}^2(t)&\ldots&\xi_{X,R_2^X}^2(t)&\ldots& 0&\ldots&0\\
 \vdots&\ddots&\vdots&\vdots&\ddots&\vdots&\ldots&\vdots&\ddots&\vdots\\
 0&\ldots&0&0&\ldots&0&\ldots &\xi_{X,1}^{p_X}(t)&\ldots&\xi_{X,R_{p_X}^X}^{p_X}(t)
 \end{pmatrix}.
 \end{align*}
\subsection{The linear functional regression latent mixture model}
\label{sectmod}
 Assuming that the $n$ observed response and covariate curves $\{({\mathbfit{y}}_1,{\mathbfit{x}}_1),\ldots,({\mathbfit{y}}_n,{\mathbfit{x}}_n)\}$ are part of $K$ homogeneous groups, we define  a latent variable ${{\mathbfit{Z}}_i}=(Z_{i1},\ldots,Z_{iK})^\top$ such that $Z_{ik}=1$ if the observation $({\mathbfit{y}_i,\mathbfit{x}}_i)$ belongs to the cluster $k$ and $Z_{ik}=0$ otherwise.  As in \cite{AntonSmith:2025}, given that $Z_{ik}=1$, $k\in\{1,\ldots,K\}$, the observations come from the following model:
 \begin{equation}
 \mathbfit{Y}_i(t)=\boldsymbol{\beta}_{0}^k(t)+\int_{\mathcal{T}_X}\boldsymbol{\beta}^k(t,s)\mathbfit{X}_i(s)ds+\mathbfit{E}^k(t), \quad t\in \mathcal{T}_Y, \quad i=1,\ldots, n. \label{reg1}
 \end{equation}
 Here $\mathbfit{E}^k(t)=(E_{1}^k(t),\ldots, E_{p_Y}^k(t))^T$ is the random error process which is uncorrelated with $\mathbfit{X}_i(s)$ for any $(s,t)\in \mathcal{T}_X \times \mathcal{T}_Y$.
 
For the random error, the regression coefficients $\boldsymbol{\beta}_{0}^k(t)=(\beta_{0,1}^k(t),\ldots, \beta_{0, p_Y}^k(t))^T$, and the $p_Y\times p_X$ matrix $\boldsymbol{\beta}^k(t,s)=\left(\beta_{lj}^k(t,s)\right)_{\substack{l=1,\ldots,p_Y\\j=1,\ldots,p_X}}$ we consider the expansions \cite[Chapter 11.3]{RamsaySilverman:2006}:    
 \begin{align}
 E_{l}^k(t)&=\sum_{r=1}^{R_l^Y}\epsilon_{0,l}^{k,r}\xi_{Y,r}^l(t), \quad l=1,\ldots,p_Y,\label{erfo}\\
 \beta_{0,l}^k(t)&=\sum_{r=1}^{R_l^Y}\Gamma_{0,l}^{k,r}\xi_{Y,r}^l(t), \quad l=1,\ldots,p_Y,\label{al1}\\
\beta_{lj}^k(t,s)&= \sum_{r_1=1}^{R_l^Y}\sum_{r_2=1}^{R_j^X}\Gamma^{k,r_1r_2}_{lj}\xi_{Y,r_1}^l(t)\xi_{X,r_2}^j(s),\quad l=1,\ldots,p_Y,\quad j=1,\ldots,p_X.\label{al2}
 \end{align}
 Notice that we have 
 \begin{equation}
 \boldsymbol{\beta}^k(t,s)=\boldsymbol{\xi}_Y(t)\boldsymbol{\Gamma}^k \boldsymbol{\xi}_X(s)^\top,\quad \boldsymbol{\beta}_{0}^k(t)=\boldsymbol{\xi}_Y(t)\boldsymbol{\Gamma}_0^k,\label{eqig}
 \end{equation}
  where $\boldsymbol{\Gamma}_0^k=(\Gamma_{0,1}^{k,1},\ldots,\Gamma_{0,1}^{k,R_1^Y},\ldots,\Gamma_{0,p_Y}^{k,1},\ldots,\Gamma_{0,p_Y}^{k,R_{p_Y}^Y})^\top$ $\in \mathbb{R}^{R^Y}$ and
 \begin{equation*}
 \boldsymbol{\Gamma}^k=\begin{pmatrix}
\Gamma_{11}^{k,11}&\ldots&\Gamma^{k,1R_1^X}_{11}&\Gamma_{12}^{k,11}&\ldots&\Gamma^{k,1R_2^X}_{12}&\ldots& \Gamma^{k,11}_{1p_X}&\ldots&\Gamma^{k,1R_{p_X}^X}_{1p_X}\\
 \vdots&\ddots&\vdots&\vdots&\ddots&\vdots&\ldots&\vdots&\ddots&\vdots\\
 \Gamma_{11}^{k,R_1^Y1}&\ldots&\Gamma^{k,R_1^YR_1^X}_{11}&\Gamma_{12}^{k,R_1^Y1}&\ldots&\Gamma^{k,R_1^YR_2^X}_{12}&\ldots& \Gamma^{k,R_1^Y1}_{1p_X}&\ldots&\Gamma^{k,R_1^YR_{p_X}^X}_{1p_X}\\
  \vdots&\ddots&\vdots&\vdots&\ddots&\vdots&\ldots&\vdots&\ddots&\vdots\\
\Gamma_{p_Y1}^{k,11}&\ldots&\Gamma^{k,1R_1^X}_{p_Y1}&\Gamma_{p_Y2}^{k,11}&\ldots&\Gamma^{k,1R_2^X}_{p_Y2}&\ldots& \Gamma^{k,11}_{p_Yp_X}&\ldots&\Gamma^{k,1R_{p_X}^X}_{p_Yp_X}\\
 \vdots&\ddots&\vdots&\vdots&\ddots&\vdots&\ldots&\vdots&\ddots&\vdots\\
 \Gamma_{p_Y1}^{k,R_{p_Y}^Y1}&\ldots&\Gamma^{k,R_{p_Y}^YR_1^X}_{p_Y1}&\Gamma_{p_Y2}^{k,R_{p_Y}^Y1}&\ldots&\Gamma^{k,R_{p_Y}^YR_2^X}_{p_Y2}&\ldots& \Gamma^{k,R_{p_Y}^Y1}_{p_Yp_X}&\ldots&\Gamma^{k,R_{p_Y}^YR_{p_X}^X}_{p_Yp_X}
 \end{pmatrix}.
 \end{equation*}

Let  $R^X:=\sum_{j=1}^{p_X} R_j^X$, $R^Y:=\sum_{l=1}^{p_Y} R_l^Y$, and ${\mathbfit{W}_X}$ be the symmetric block-diagonal $R^X\times R^X$ matrix of inner products between the basis functions:
$$
{\mathbfit{W}_X}=\int_{\mathcal{T}_X}\boldsymbol{\xi}_X(s)^\top\boldsymbol{\xi}_X(s)ds, 
$$
Using  \eqref{eqsecond}-\eqref{eqig}, for any $i=1,\ldots, n$ for which $Z_{ik}=1$  we get , 
\begin{align*}
&\mathbfit{Y}_i^\top(t)={\mathbfit{c}_{Y,i}}^\top\boldsymbol{\xi}_Y^\top (t)=\boldsymbol{\beta}_{0}^k(t)^\top+\int_{\mathcal{T}_X}\mathbfit{X}_i^\top (s)\boldsymbol{\beta}^k(t,s)^\top ds+\mathbfit{E}^k(t)^\top\\
&=\left(\left(\boldsymbol{\Gamma}_0^k\right)^\top+{\mathbfit{c}_{X,i}}^\top{\mathbfit{W}_X}\left(\boldsymbol{\Gamma}^k\right)^\top +\left(\boldsymbol{\epsilon}_{0}^k\right)^\top\right)
\boldsymbol{\xi}_Y^\top(t),
\end{align*}
for any $t\in \mathcal{T}_Y$. Here  we denote $\boldsymbol{\epsilon}_0^k=(\epsilon_{0,1}^{k,1},\ldots,\epsilon_{0,1}^{k,R_1^Y},\ldots,$ $\epsilon_{0,p_Y}^{k,1},\ldots,\epsilon_{0,p_Y}^{k,R_{p_Y}^Y})^\top\in \mathbb{R}^{R^Y}$, and ${\mathbfit{c}}_{X,i}$, ${\mathbfit{c}}_{Y,i}$ are column vectors formed with the coefficients  in the $i$th row of the matrices ${\mathbfit{C}_X}$ and  ${\mathbfit{C}_Y}$ respectively.
Thus, given that $Z_{ik}=1$, we obtain the following model for  $\mathbfit{c}_{Y,i}$:
\begin{equation}
\mathbfit{c}_{Y,i}=\boldsymbol{\Gamma}_0^k+\boldsymbol{\Gamma}^k\mathbfit{W}_X{\mathbfit{c}_{X,i}} +\boldsymbol{\epsilon}_{0}^k.\label{coeffsmat}
\end{equation}
The same model was considered in \cite{AntonSmith:2025}, but instead of assuming multivariate normal distributions for the coefficients, here we consider three multivariate skewed distributions.
 
As for funHDDC in \cite{SchmutzJacquesBouveyronChezeMartin:2020} we use MFPCA to represent the stochastic process $\mathbfit{X}$ associated with the $k$th cluster, $k\in\{1,\ldots,K\}$, in a lower dimensional subspace $\mathbb{E}^k[0,\mathcal{T}_X]\subset L^2[0,\mathcal{T}_X]$ with dimension $d_k\le R^X$. The MFPCA scores are obtained directly from a principal component analysis of the coefficients ${\mathbfit{C_X}}$ with a metric based on the inner products between the basis functions included in ${\mathbfit{W_X}}$. We split the orthogonal $R^X \times R^X$ matrix ${\mathbfit{Q}}_k=(q_{krj})_{r,j=1,\ldots,R^X}$  containing the coefficients of the eigenfunctions expressed in the initial basis ${\boldsymbol{\xi}}$ as ${\mathbfit Q}_k=[{\mathbfit{U}}_k,{\mathbfit{V}}_k]$ such that ${\mathbfit{U}}_k$ is of size $R^X\times d_k$, ${\mathbfit{V}}_k$ is of size $R^X\times (R^X- d_k)$ and we have
\begin{equation*}
{\mathbfit Q}_k^\top {\mathbfit Q}_k={\mathbfit I}_{R^X},\quad {\mathbfit{U}}_k^\top {\mathbfit{U}}_k={\mathbfit I}_{d_k},\quad {\mathbfit{V}}_k^\top {\mathbfit{V}}_k={\mathbfit I}_{R^X-d_k},\quad {\mathbfit{U}}_k^\top {\mathbfit{V}}_k=\mathbf{0}.
\end{equation*}

We can make distribution assumptions on the scores \citep{DelaigleHall:2010}, such that for the $k$th cluster $\mathbfit{c}_{X,i}$ arises from  the VG, the ST, or the NIG distribution presented in Section \ref{sectskew}, and the density $f_k(\mathbfit{c}_{X,i} \mid \boldsymbol{\theta}_{X,k})$ is given by equation \eqref{pdfVG}, \eqref{pdfST} or \eqref{pdfNIG}, respectively, where $\boldsymbol{\theta}_{X,k}$ is the set formed with the parameters.  Using \eqref{skewm} we introduce the latent random variable $W_{X,i}>0$, $i=1,\ldots, n$ such that, independently for $i=1,\ldots, n$, depending on the skewed distribution considered we have 
 \begin{itemize}
 \item $W_{X,i}\mid Z_{ik}=1\sim G(\psi_{X,k},\psi_{X,k})$, $\psi_{X,k}>0$, and $\mathbfit{c}_{X,i}\mid Z_{ik}=1\sim VG_{R^X}(\boldsymbol{\mu}_{X,k},\boldsymbol{\alpha}_{X,k},\boldsymbol{\Sigma}_{X,k}$, $\psi_{X,k})$;
 \item $W_{X,i}\mid Z_{ik}=1\sim IG(\nu_{X,k}/2,\nu_{X,k}/2)$, $\nu_{X,k}>0$, and  $\mathbfit{c}_{X,i}\mid Z_{ik}=1\sim ST_{R^X}(\boldsymbol{\mu}_{X,k},\boldsymbol{\alpha}_{X,k}$, $\boldsymbol{\Sigma}_{X,k},\nu_{X,k})$;
 \item $W_{X,i}\mid Z_{ik}=1\sim IN(1,\kappa_{X,k}/2)$, $\kappa_{X,k}>0$, and $\mathbfit{c}_{X,i}\mid Z_{ik}=1\sim NIG_{R^X}(\boldsymbol{\mu}_{X,k},\boldsymbol{\alpha}_{X,k}$, $\boldsymbol{\Sigma}_{X,k},\kappa_{X,k})$.
 \end{itemize}
 Thus 
 \begin{equation}
 \mathbfit{c}_{X,i}\mid W_{X,i}=w_{i,X},Z_{ik}=1 \sim N({\boldsymbol{\mu}}_{X,k}+w_{i,X}\boldsymbol{\alpha}_{X,k},w_{i,X}\boldsymbol{\Sigma}_{X,k}),\label{distc}
 \end{equation}
 where 
\begin{equation}
{\mathbfit{D}}_k={\mathbfit{Q}}_k^\top {\mathbfit{W}_X}^{1/2}\boldsymbol\Sigma_{X,k}{\mathbfit{W}_X}^{1/2}{\mathbfit{Q}}_k=\text{diag}(a_{k1},\ldots, a_{kd_k}, b_k,\ldots, b_k),\label{dis5}
\end{equation}
with $a_{k1}>a_{k2}>\cdots >a_{kd_k}> b_k$. 
Let $\phi({\mathbfit{c}}_{X,i};\boldsymbol{\mu}_{X,k}+w_{i,X}\boldsymbol{\alpha}_{X,k},w_i\boldsymbol{\Sigma}_{X,k})$ denotes the density for the $R_X-$variate normal distribution $N(\boldsymbol{\mu}_{X,k}+w_{i,X}\boldsymbol{\alpha}_{X,k},w_{i,X}\boldsymbol{\Sigma}_{X,k})$ 
\begin{align}
&\phi({\mathbfit{c}}_{X,i};\boldsymbol{\mu}_{X,k}+w_{i,X}\boldsymbol{\alpha}_{X,k},w_i\boldsymbol{\Sigma}_{X,k})=(2\pi)^{-R^X/2}w_{i,X}^{-1/2}\mid \boldsymbol{\Sigma}_{X,k}\mid ^{-1/2}\notag\\
&\exp\biggl(-\frac{1}{2w_{i,X}}({\mathbfit{c}}_{X,i}-\boldsymbol{\mu}_{X,k}-w_{i,X}\boldsymbol{\alpha}_{X,k})^\top\boldsymbol{\Sigma}_{X,k}^{-1}({\mathbfit{c}}_{X,i}-\boldsymbol{\mu}_{X,k}-w_{i,X}\boldsymbol{\alpha}_{X,k})\biggl).\label{densN}
\end{align} 
Here  $\mid \boldsymbol{\Sigma}_{X,k}\mid$ denotes the determinant of $\boldsymbol{\Sigma}_{X,k}$.

Similarly, we assume that for the $k$th cluster, the conditional density $g_k(\mathbfit{c}_{Y,i} \mid \mathbfit{c}_{X,i},\boldsymbol{\theta}_{Y,k})$ of $\mathbfit{c}_{Y,i}$ given  $\mathbfit{c}_{X,i}$ and $Z_{ik}=1$ corresponds to one of the densities given in formulas \eqref{pdfVG}, \eqref{pdfST}, or \eqref{pdfNIG}. Thus,  using \eqref{skewm} we introduce the latent random variable $W_{Y,i}>0$, $i=1,\ldots, n$ such that,  independently for $i=1,\ldots, n$, according to the distribution considered, we have
 \begin{itemize}
 \item $W_{Y,i}\mid Z_{ik}=1, \mathbfit{c}_{X,i} \sim G(\psi_{Y,k},\psi_{Y,k})$, $\psi_{Y,k}>0$ and $\mathbfit{c}_{Y,i}\mid Z_{ik}=1, \mathbfit{c}_{X,i}\sim VG_{R^Y}(\boldsymbol{\mu}_{Y,k},\boldsymbol{\alpha}_{Y,k}$, $\boldsymbol{\Sigma}_{Y,k},\psi_{Y,k})$;
 \item $W_{Y,i}\mid Z_{ik}=1, \mathbfit{c}_{X,i}\sim IG(\nu_{Y,k}/2,\nu_{Y,k}/2)$, $\nu_{Y,k}>0$ and $\mathbfit{c}_{Y,i}\mid Z_{ik}=1, \mathbfit{c}_{X,i}\sim ST_{R^Y}(\boldsymbol{\mu}_{Y,k},\boldsymbol{\alpha}_{Y,k},\boldsymbol{\Sigma}_{Y,k},\nu_{Y,k})$;
 \item $W_{Y,i}\mid Z_{ik}=1, \mathbfit{c}_{X,i}\sim IN(1,\kappa_{Y,k}/2)$, $\kappa_{Y,k}>0$ and $\mathbfit{c}_{Y,i}\mid Z_{ik}=1, \mathbfit{c}_{X,i}\sim NIG_{R^Y}(\boldsymbol{\mu}_{Y,k},\boldsymbol{\alpha}_{Y,k}$, $\boldsymbol{\Sigma}_{Y,k},\kappa_{Y,k})$.
 \end{itemize}

Given that $Z_{ik}=1$, from \eqref{coeffsmat} we have for $\mathbfit{c}_{Y,i}$:
\begin{equation*}
\mathbfit{c}_{Y,i}=\boldsymbol{\Gamma}^k_{*}{\mathbfit{c}_{X,i}^{*}}+\boldsymbol{\epsilon}_{0}^k,
\end{equation*}
where $\mathbfit{c}_{X,i}^{*}=\left( \mathbfit{W}_X\mathbfit{c}_{X,i} ,1 \right)^\top$ and $\boldsymbol{\Gamma}^k_{*}$ is the $R^Y\times (R^{X}+1) $ matrix $\boldsymbol{\Gamma}^k_{*}=(\boldsymbol{\Gamma}^k,\boldsymbol{\Gamma}_0^k)$. Hence
 \begin{align}
& \mathbfit{c}_{Y,i}\mid W_{Y,i}=w_{i,Y}, Z_{ik}=1, \mathbfit{c}_{X,i} \sim N({\boldsymbol{\mu}}_{Y,k}+w_{i,Y}\boldsymbol{\alpha}_{Y,k},w_{i,Y}\boldsymbol{\Sigma}_{Y,k}),\quad \boldsymbol{\mu}_{Y,k}=\boldsymbol{\Gamma}^k_{*}{\mathbfit{c}_{X,i}^{*}} \label{distyy},
\end{align}

Thus the joint distribution of the coefficients $(\mathbfit{c}_{Y,i}\mathbfit{c}_{X,i})$,  $i=1,\ldots, n$ arise from a parametric mixture distribution
\begin{align}
p(\mathbfit{c}_{Y,i},\mathbfit{c}_{X,i};\boldsymbol{\theta})&=\sum_{k=1}^K\pi_k p_k(\mathbfit{c}_{Y,i},\mathbfit{c}_{X,i} \mid \boldsymbol{\theta}_k), \quad \sum_{k=1}^K\pi_k=1, \label{dfun0}\\
p_k(\mathbfit{c}_{Y,i},\mathbfit{c}_{X,i} \mid \boldsymbol{\theta}_{k})&=f_k(\mathbfit{c}_{X,i} \mid \boldsymbol{\theta}_{X,k})g_k(\mathbfit{c}_{Y,i} \mid \mathbfit{c}_{X,i},\boldsymbol{\theta}_{Y,k}),\label{dens}
\end{align}
where $\pi_k\in (0,1]$ are the mixing proportions, $\boldsymbol{\theta}_k$ are the parameters for the $k$th cluster and $\boldsymbol{\theta}=\bigcup_{k=1}^k (\boldsymbol{\theta}_{X,k} \cup\boldsymbol{\theta}_{Y,k} \cup\{\pi_k\})$,  is the set formed with the parameters. Notice that $f_k(\mathbfit{c}_{X,i} \mid \boldsymbol{\theta}_{X,k})$ and  $g_k(\mathbfit{c}_{Y,i} \mid \mathbfit{c}_{X,i},\boldsymbol{\theta}_{Y,k})$ are not necessary of the same type, so we have 9 combinations of skewed distributions.  
 
We refer to this model as a FLM[$a_{kj},b_k,{\mathbfit{Q}}_k,d_k$] - VVV model. As in Table 1 in \cite{ BouveyronJacques:2011} we consider five more  parsimonious sub-models  of the FLM[$a_{kj},b_k,{\mathbfit{Q}}_k,d_k$] (functional latent mixture) for which the parameters $b_k$ are common between the clusters and/or the first $d_k$ diagonal elements of $\mathbfit{D}_k$ are common within each cluster or  are common within each cluster and between the clusters. 

We consider parsimony also for the matrices $\boldsymbol{\Sigma}_{Y,k}$ by  constraining their eigen-decomposition $\boldsymbol{\Sigma}_{Y,k} = \lambda_k\boldsymbol{\Xi}_k\boldsymbol{\Upsilon}_k\boldsymbol{\Xi}_k^\top$, where $\boldsymbol{\Upsilon}_k$  is a diagonal matrix
with entries (sorted in decreasing order) proportional to the eigenvalues of
$\boldsymbol{\Sigma}_{Y,k} $ with the constraint $\mid\boldsymbol{\Upsilon}_k\mid = 1$,  $\boldsymbol{\Xi}_k$ is a $R^Y \times R^Y$ orthogonal matrix of the
eigenvectors (ordered according to the eigenvalues) of $\boldsymbol{\Sigma}_{Y,k} $, and $\lambda_k=\mid\boldsymbol{\Sigma}_{Y,k} \mid^{1/R^Y}$ is a constant, $k = 1, \ldots K$. Following \cite{CeleuxGovaert:1995} we get 13 more parsimonious models by constraining some of these parameters to be equal between groups. Overall we obtain $6 \times 14=84$ parsimonious models.

We know investigate the identifiability of the model \eqref{dfun0}-\eqref{dens} based on a pair of skewed distributions VG, ST or NIG. In \cite{GallaugherTomarchioMcNicholasPunzo:2022} it is shown that the VG, ST and NIG distributions are nested in the generalized hyperbolic (GH) distribution. Moreover,  Theorem 3.1 in \cite{GallaugherTomarchioMcNicholasPunzo:2022} shows the identifiability of the GH-GH cluster weighted model if the parameters  $\boldsymbol{\theta}_{Y,k}$ are pairwise distinct and the densities $f(x| \boldsymbol{\theta}_{X,k})$ are not degenerate,
$k = 1, \ldots, K$.  In the space of the coefficients this result directly implies the identifiability  under similar conditions  of the model \eqref{dfun0}-\eqref{dens} based on a pair of skewed distributions VG, ST or NIG.



\section{Parameter inference for the FLM[$a_{kj},b_k,{\mathbfit{Q}}_k,d_k$] - VVV model}
\label{section3}
The clusters' labels ${\mathbfit{Z}_i}$ and the values $W_{X,i}$, $W_{Y,i}$ are not observed, so to estimate the parameters we use the  expectation-maximization (EM) algorithm \citep{DempsterLairdRubin:1977}. Each iteration of the EM algorithm has two steps, the expectation (E) and the maximization (M) steps. In the  E step the conditional expectation of the complete data log-likelihood $l_c(\boldsymbol{\theta})$ is computed based on the current estimates of the parameters $\boldsymbol{\theta}$, where the complete data consists of $\{\mathbfit{c}_{Y,i}, \mathbfit{c}_{X,i}, z_{ik},w_{i,X}, w_{i,Y}$ $ i=1,\ldots, n, k=1,\ldots,K\}$. In the M step  the estimates of $\boldsymbol{\theta}$ are updated with the values that maximize the expected complete log-likelihood. The formulas for $l_c(\boldsymbol{\theta})$ is included in Proposition \ref{proplik}  in Appendix \ref{secA1}. 

Next we present the EM algorithm for the most general model   FLM[$a_{kj},b_k,{\mathbfit{Q}}_k,d_k$] - VVV model. For the other parsimonious models the updates in the M-step are different only for the covariance matrix $\boldsymbol{\Sigma}_{Y,k}^{(m)}$ and $a_{kj}^{(m)}$, $b_k^{(m)}$, $k=1,\ldots, K, j=1,\ldots, d_k$.
For the simplified FLM models and for the covariance matrix $\boldsymbol{\Sigma}_{Y,k}^{(m)}$ with constraint eigen-decomposition, the updates are similar with those of the Gaussian parsimonious
clustering models in  \cite{ BouveyronJacques:2011} and  \cite{CeleuxGovaert:1995}, respectively. 
\subsubsection{The E-step}
 We calculate $E[l_c(\boldsymbol{\theta}^{(m-1)})\mid \mathbfit{c}_{Y,1},\mathbfit{c}_{X,1}\ldots,\mathbfit{c}_{Y,n},\mathbfit{c}_{X,n},\boldsymbol{\theta}^{(m-1)}]$, given the current values of the parameters $\boldsymbol{\theta}^{(m-1)}$.  This reduces to the calculation of 
\begin{align*}
&w_{ik,X}^{(m)}:=E[W_{X,i}\mid Z_{ik}=1,\mathbfit{c}_{X,1}\ldots,\mathbfit{c}_{X,n},\boldsymbol{\theta}^{(m-1)}],\\
&wi_{ik,X}^{(m)}=E[1/W_{X,i}\mid Z_{ik}=1,\mathbfit{c}_{X,1}\ldots,\mathbfit{c}_{X,n},\boldsymbol{\theta}^{(m-1)}],\\
&lw_{ik,X}^{(m)}= E[\log(W_{X,i})\mid Z_{ik}=1,\mathbfit{c}_{X,1}\ldots,\mathbfit{c}_{X,n},\boldsymbol{\theta}^{(m-1)}],\\
&w_{ik,Y}^{(m)}:=E[W_{Y,i}\mid Z_{ik}=1,\mathbfit{c}_{Y,1},\mathbfit{c}_{X,1}\ldots,\mathbfit{c}_{Y,n},\mathbfit{c}_{X,n}\boldsymbol{\theta}^{(m-1)}],\\
&wi_{ik,Y}^{(m)}=E[1/W_{Y,i}\mid Z_{ik}=1,\mathbfit{c}_{Y,1},\mathbfit{c}_{X,1}\ldots,\mathbfit{c}_{Y,n},\mathbfit{c}_{X,n},\boldsymbol{\theta}^{(m-1)}],\\
&lw_{ik,Y}^{(m)}=E[\log(W_{Y,i})\mid Z_{ik}=1,\mathbfit{c}_{Y,1},\mathbfit{c}_{X,1}\ldots,\mathbfit{c}_{Y,n},\mathbfit{c}_{X,n},\boldsymbol{\theta}^{(m-1)}]\\
&t_{ik}^{(m)}:=E[Z_{ik}\mid \mathbfit{c}_{Y,1},\mathbfit{c}_{X,1}\ldots,\mathbfit{c}_{Y,n},\mathbfit{c}_{X,n},\boldsymbol{\theta}^{(m-1)}]=\frac{\pi_k p_k\left(\mathbfit{c}_{Y,i},\mathbfit{c}_{X,i} \mid \boldsymbol{\theta}_k^{(m-1)}\right)}{\sum_{l=1}^K\pi_l p_l\left(\mathbfit{c}_{Y,i},\mathbfit{c}_{X,i} \mid \boldsymbol{\theta}_l^{(m-1)}\right)}\notag.
\end{align*}
We calculate  $w_{ik,X}^{(m)}$, $wi_{ik,X}^{(m)}$, $lw_{ik,X}^{(m)}$, $w_{ik,Y}^{(m)}$, $wi_{ik,Y}^{(m)}$, $lw_{ik,Y}^{(m)}$, $i=1,\ldots, n$, $k=1,\ldots, K$ using formulas \eqref{es1}-\eqref{es3}.

Notice that the probability densities functions in formulas  \eqref{pdfVG}, \eqref{pdfST}, and \eqref{pdfNIG} can be written using  a single formula as
\begin{align}
&f_{SKW}(\mathbfit{v};\boldsymbol{\mu,\alpha,\Sigma},p_1,p_2,p_3,p_4)=\exp\biggl(\left(\mathbfit{v}-\boldsymbol{\mu}\right)^\top\boldsymbol{\Sigma}^{-1}\boldsymbol{\alpha}+p_3\log\left(\delta(\mathbfit{v};\boldsymbol{\mu}, \boldsymbol{\Sigma})+p_1\right)\notag\\
&-p_3\log\left(\rho(\boldsymbol{\alpha,\Sigma})+p_2\right)+\log\left( K_{2p_3}\left(\sqrt{\left(\rho(\boldsymbol{\alpha,\Sigma})+p_2\right)\left(\delta(\mathbfit{v};\boldsymbol{\mu, \Sigma})+p_1\right)}\right)\right)\notag\\
&-\frac{d}{2}\log\left(2\pi\right)-\frac{1}{2}\log\left(|\boldsymbol{\Sigma}|\right)+p_4\biggl),\label{pdfUNIF}
\end{align}
where 
\begin{itemize}
\item[VG:]  $p_1=0$, $p_2=2\psi$, $p_3=(\psi-d/2)/2$, $p_4=\psi\log(\psi)-\log\left(\Gamma(\psi)\right)+\log (2)$;
\item[ST:]  $p_1=\nu$, $p_2=0$, $p_3=-(\nu+d)/4$, $p_4=\frac{\nu}{2}\log\left(\frac{\nu}{2}\right)-\log\left(\Gamma\left(\frac{\nu}{2}\right)\right)+\log (2)$;
\item[NIG:] $p_1=1$, $p_2=\kappa^2$, $p_3=-(1+d)/4$, $p_4=\kappa+\frac{1}{2}\log\left(\frac{2}{\pi}\right)$;
\end{itemize}
 \begin{proposition}
 \label{propEM}
Let us denote
\begin{align}
&\delta_{X,k}(\mathbfit{c}_{X,i})=\biggl(\sum_{l=1}^{d_k}\frac{\mathbfit{q}_{kl}^\top \mathbfit{W}_X^{1/2}(\mathbfit{c}_{X,i}-\boldsymbol{\mu}_{X,k})(\mathbfit{c}_{X,i}-\boldsymbol{\mu}_{X,k})^\top \mathbfit{W}_X^{1/2}\mathbfit{q}_{kl}}{a_{kl}}\nonumber\\
&+\sum_{l=d_k+1}^{R}\frac{\mathbfit{q}_{kl}^\top \mathbfit{W}_X^{1/2}(\mathbfit{c}_{X,i}-\boldsymbol{\mu}_{X,k})(\mathbfit{c}_{X,i}-\boldsymbol{\mu}_{X,k})^\top\mathbfit{W}_X^{1/2}\mathbfit{q}_{kl}}{b_{k}}\biggl)\label{newdelta}\\
&\rho_{X,k}=\biggl(\sum_{l=1}^{d_k}\frac{\mathbfit{q}_{kl}^\top \mathbfit{W}_X^{1/2}\boldsymbol{\alpha}_{X,k}\boldsymbol{\alpha}_{X,k}^\top \mathbfit{W}_X^{1/2}\mathbfit{q}_{kl}}{a_{kl}}+\sum_{l=d_k+1}^{R}\frac{\mathbfit{q}_{kl}^\top \mathbfit{W}_X^{1/2}\boldsymbol{\alpha}_{X,k}\boldsymbol{\alpha}_{X,k}^\top\mathbfit{W}_X^{1/2}\mathbfit{q}_{kl}}{b_{k}}\biggl)\label{newro}\\
&\delta_{\alpha,k}(\mathbfit{c}_{X,i})=(\mathbfit{c}_{X,i}-\boldsymbol{\mu}_{X,k})^\top \boldsymbol{\Sigma}_{X,k}^{-1}\boldsymbol{\alpha}_{X,k}=\biggl(\sum_{l=1}^{d_k}\frac{\mathbfit{q}_{kl}^\top \mathbfit{W}_X^{1/2}\boldsymbol{\alpha}_{X,k}(\mathbfit{c}_{X,i}-\boldsymbol{\mu}_{X,k})^\top \mathbfit{W}_X^{1/2}\mathbfit{q}_{kl}}{a_{kl}}\nonumber\\
&+\sum_{l=d_k+1}^{R}\frac{\mathbfit{q}_{kl}^\top \mathbfit{W}_X^{1/2}\boldsymbol{\alpha}_{X,k}(\mathbfit{c}_{X,i}-\boldsymbol{\mu}_{X,k})^\top\mathbfit{W}_X^{1/2}\mathbfit{q}_{kl}}{b_{k}}\biggl)\label{newdeltaalf}\\
&H_k(\mathbfit{c}_{Y,i},\mathbfit{c}_{X,i}\mid \boldsymbol{\theta}_k)=\log(\pi_k)-\frac{1}{2}\biggl(\sum_{j=1}^{d_k} \log (a_{kj})+(R_X-d_k)\log (b_k)+\log(\mid\boldsymbol{\Sigma}_{Y,k}\mid)\biggl)\notag\\
&+\delta_{\alpha,k}(\mathbfit{c}_{X,i})+p_{X,3,k}\log\left(\delta_{X,k}(\mathbfit{c}_{X,i})+p_{X,1,k}\right)-p_{X,3,k}\log\left(\rho_{X,k}+p_{X,2,k}\right)+p_{X,4,k}\notag\\
&+\log\left( K_{2p_{X,3,k}}\left(\sqrt{\left(\rho_{X,k}+p_{X,2,k}\right)\left(\delta_{X,k}(\mathbfit{c}_{X,i})+p_{X,1,k}\right)}\right)\right)\notag\\
&+\left(\mathbfit{c}_{Y,i}-\boldsymbol{\Gamma}^k_{*}{\mathbfit{c}_{X,i}^{*}}\right)^\top\boldsymbol{\Sigma}_{Y,k}^{-1}\boldsymbol{\alpha}_{Y,k}+p_{Y,3,k}\log\left(\delta(\mathbfit{c}_{Y,i};\boldsymbol{\Gamma}^k_{*}{\mathbfit{c}_{X,i}^{*}}, \boldsymbol{\Sigma}_{Y,k})+p_{Y,1,k}\right)\notag\\
&-p_{Y,3,k}\log\left(\rho(\boldsymbol{\alpha}_{Y,k},\boldsymbol{\Sigma}_{Y,k})+p_{Y,2,k}\right)+p_{Y,4,k}\notag\\
&+\log\left( K_{2p_{Y,3,k}}\left(\sqrt{\left(\rho(\boldsymbol{\alpha}_{Y,k},\boldsymbol{\Sigma}_{Y,k})+p_{Y,2,k}\right)\left(\delta(\mathbfit{c}_{Y,i};\boldsymbol{\Gamma}^k_{*}{\mathbfit{c}_{X,i}^{*}}, \boldsymbol{\Sigma}_{Y,k})+p_{Y,1,k}\right)}\right)\right).\label{Hk}
\end{align}
We have
\begin{align}
&t_{ik}^{(m)}:=E[Z_{ik}\mid \mathbfit{c}_{Y,1},\mathbfit{c}_{X,1}\ldots,\mathbfit{c}_{Y,n},\mathbfit{c}_{X,n},\boldsymbol{\theta}^{(m-1)}]=\frac{\pi_k p_k\left(\mathbfit{c}_{Y,i},\mathbfit{c}_{X,i} \mid \boldsymbol{\theta}_k^{(m-1)}\right)}{\sum_{l=1}^K\pi_l p_l\left(\mathbfit{c}_{Y,i},\mathbfit{c}_{X,i} \mid \boldsymbol{\theta}_l^{(m-1)}\right)}\notag\\
&=\frac{1}{\sum_{l=1}^K\exp \left(\left(H_l\left(\mathbfit{c}_{Y,i},\mathbfit{c}_{X,i}\mid \boldsymbol{\theta}_k^{(m-1)}\right)-H_k\left(\mathbfit{c}_{Y,i},\mathbfit{c}_{X,i}\mid \boldsymbol{\theta}_l^{(m-1)}\right)\right)\right)},\label{newt}
\end{align}
\end{proposition}
\begin{proof}
 The proof is included in  Appendix \ref{secA2}.
 \end{proof}

Based on the current values of the parameters $\boldsymbol{\theta}^{(m-1)}$ the log-likelihood is given by
\begin{align*}
\L^{(m-1)}&=\log\left(\prod_{i=1}^n p\left(\mathbfit{c}_{Y,i},\mathbfit{c}_{X,i};\boldsymbol{\theta}^{(m-1)}\right) \right)\\
&=\sum_{i=1}^n\log\left(\sum_{k=1}^K\pi_k^{(m-1)} p_k\left(\mathbfit{c}_{Y,i},\mathbfit{c}_{X,i} \mid \boldsymbol{\theta}_k^{(m-1)}\right)\right)
\end{align*}
\subsubsection{The M-step}
We update the parameters  by maximizing the conditional expectation of the complete data log likelihood  $Q(\boldsymbol{\theta}\mid \boldsymbol{\theta}^{(m-1)}):=E[\log(l_c(\boldsymbol{\theta}^{(m-1)}))\mid \mathbfit{c}_{Y,1},\mathbfit{c}_{X,1},\ldots,\mathbfit{c}_{Y,n},\mathbfit{c}_{X,n},\boldsymbol{\theta}^{(m-1)}]$.
\begin{proposition}
\label{propM}
Let 
\begin{align}
&\mathbfit{S}_{X,k}^{(m)}=\frac{\sum_{i=1}^n t_{ik}^{(m)}wi_{ik,X}^{(m)}(\mathbfit{c}_{X,i}-\boldsymbol{\mu}_{X,k}^{(m)})(\mathbfit{c}_{X,i}-\boldsymbol{\mu}_{X,k}^{(m)})^\top}{n_k^{(m)}}+\frac{\sum_{i=1}^n t_{ik}^{(m)}w_{ik,X}^{(m)}\boldsymbol{\alpha}_{X,k}^{(m)}(\boldsymbol{\alpha}_{X,k}^{(m)})^\top}{n_k^{(m)}}\notag\\
&-\frac{\sum_{i=1}^n t_{ik}^{(m)}\left((\mathbfit{c}_{X,i}-\boldsymbol{\mu}_{X,k}^{(m)})\boldsymbol{\alpha}_{X,k}^\top+\boldsymbol{\alpha}_{X,k}^{(m)}(\mathbfit{c}_{X,i}-\boldsymbol{\mu}_{X,k}^{(m)})^\top\right)}{n_k^{(m)}}, \quad n_k^{(m)}=\sum_{i=1}^n t_{ik}^{(m)},\label{sig}\\
&\bar{w}_{k,X}^{(m)}=\frac{\sum_{i=1}^n t_{ik}^{(m)}w_{ik,X}^{(m)}}{n_{k}^{(m)}},~\bar{wi}_{k,X}^{(m)}=\frac{\sum_{i=1}^n t_{ik}^{(m)}wi_{ik,X}^{(m)}}{n_{k}^{(m)}},~\bar{lw}_{k,X}^{(m)}=\frac{\sum_{i=1}^n t_{ik}^{(m)}lw_{ik,X}^{(m)}}{n_{k}^{(m)}},\label{extradefx}\\
&\bar{w}_{k,Y}^{(m)}=\frac{\sum_{i=1}^n t_{ik}^{(m)}w_{ik,Y}^{(m)}}{n_{k}^{(m)}},~\bar{wi}_{k,Y}^{(m)}=\frac{\sum_{i=1}^n t_{ik}^{(m)}wi_{ik,Y}^{(m)}}{n_{k}^{(m)}},~\bar{lw}_{k,Y}^{(m)}=\frac{\sum_{i=1}^n t_{ik}^{(m)}lw_{ik,Y}^{(m)}}{n_{k}^{(m)}}.\label{extradefy}
\end{align}
For the model FLM[$a_{kj},b_k,{\mathbfit{Q}}_k,d_k$]-VVV  we have the following updates for the parameters
\begin{align}
&\pi_k^{(m)}=\frac{\sum_{i=1}^n t_{ik}^{(m)}}{n}=\frac{n_k^{(m)}}{n},\quad k=1,\ldots, K, \label{tpi}\\
&{\boldsymbol{\mu}}_{X,k}^{(m)}=\frac{\sum_{i=1}^n t_{ik}^{(m)}\mathbfit{c}_{X,i}\left(wi_{ik,X}^{(m)}\left(\sum_{i=1}^n t_{ik}^{(m)}w_{ik,X}^{(m)}\right)-n_{k}^{(m)}\right)}{\left(\sum_{i=1}^n t_{ik}^{(m)}w_{ik,X}^{(m)}\right)\left(\sum_{i=1}^n t_{ik}^{(m)}wi_{ik,X}^{(m)}\right)-\left(n_{k}^{(m)}\right)^2}\notag\\
&=\frac{\sum_{i=1}^n t_{ik}^{(m)}\mathbfit{c}_{X,i}\left(wi_{ik,X}^{(m)}\bar{w}_{k,X}^{(m)}-1\right)}{\bar{w}_{k,X}^{(m)}\left(\sum_{i=1}^n t_{ik}^{(m)}wi_{ik,X}^{(m)}\right)-n_{k}^{(m)}}, \quad k=1,\ldots, K, \label{tmu}\\
&{\boldsymbol{\alpha}}_{X,k}^{(m)}=\frac{\sum_{i=1}^n t_{ik}^{(m)}\mathbfit{c}_{X,i}\left(\sum_{i=1}^n t_{ik}^{(m)}wi_{ik,X}^{(m)}\right)-n_{k}^{(m)}\sum_{i=1}^n t_{ik}^{(m)}\mathbfit{c}_{X,i}wi_{ik,X}^{(m)}}{\sum_{i=1}^n t_{ik}^{(m)}w_{ik,X}^{(m)}\left(\sum_{i=1}^n t_{ik}^{(m)}wi_{ik,X}^{(m)}\right)-\left(n_{k}^{(m)}\right)^2}\notag\\
&=\frac{\sum_{i=1}^n t_{ik}^{(m)}\mathbfit{c}_{X,i}\left(\bar{wi}_{k,X}^{(m)}-wi_{k,X}^{(m)}\right)}{\bar{w}_{k,X}^{(m)}\sum_{i=1}^n t_{ik}^{(m)}wi_{ik,X}^{(m)}-n_{k}^{(m)}},\quad k=1,\ldots, K,  \label{formalpha}
\end{align}
\begin{itemize}
\item ${\mathbfit{q}}_{kj}^{(m)}$, $k=1,\ldots, K, j=1,\ldots, d_k$ are updated as the eigenfunctions associated with the $d_k$ largest  eigenvalues of $\mathbfit{W}_X^{1/2}\mathbfit{S}_{X,k}^{(m)}\mathbfit{W}_X^{1/2}$;
\item $a_{kj}^{(m)}$,  $k=1,\ldots, K, j=1,\ldots, d_k$ are updated by the $d_k$ largest  eigenvalues of $\mathbfit{W}_X^{1/2}\mathbfit{S}_{X,k}^{(m)}\mathbfit{W}_X^{1/2}$;
\item $b_k^{(m)}$, $k=1,\ldots, K$ are updated by
\begin{equation}
b_k^{(m)}=\frac{1}{R_X-d_k}\left(\text{trace}\left(\mathbfit{W}_X^{1/2}\mathbfit{S}_{X,k}^{(m)}\mathbfit{W}_X^{1/2}\right)-\sum_{j=1}^{d_k} a_{kj}^{(m)}\right).\label{tb1}
\end{equation}
\end{itemize}
\begin{align}
&(\boldsymbol{\Gamma}^k_{*})^{(m)}=\left(\sum_{i=1}^n t_{ik}^{(m)}wi_{ik,Y}^{(m)}\mathbfit{c}_{Y,i}({\mathbfit{c}_{X,i}^{*}})^\top-\frac{1}{n_{k}^{(m)}\bar{w}_{k,Y}^{(m)}}\sum_{i=1}^n t_{ik}^{(m)}\mathbfit{c}_{Y,i}\sum_{i=1}^n t_{ik}^{(m)}({\mathbfit{c}_{X,i}^{*}})^\top\right)\notag\\
&\left(\sum_{i=1}^n t_{ik}^{(m)}wi_{ik,Y}^{(m)}{\mathbfit{c}_{X,i}^{*}}({\mathbfit{c}_{X,i}^{*}})^\top-\frac{1}{n_{k}^{(m)}\bar{w}_{k,Y}^{(m)}}\sum_{i=1}^n t_{ik}^{(m)}{\mathbfit{c}_{X,i}^{*}}\sum_{i=1}^n t_{ik}^{(m)}({\mathbfit{c}_{X,i}^{*}})^\top\right)^{-1},\label{GammY}\\
&\boldsymbol{\alpha}_{Y,k}^{(m)}=\frac{1}{n_{k}^{(m)}\bar{w}_{k,Y}^{(m)}}\left(\sum_{i=1}^n t_{ik}^{(m)}\mathbfit{c}_{Y,i}-(\boldsymbol{\Gamma}^k_{*})^{(m)}\sum_{i=1}^n t_{ik}^{(m)}{\mathbfit{c}_{X,i}^{*}}\right), ~\boldsymbol{\mu}_{Y,k}^{(m)}=(\boldsymbol{\Gamma}^k_{*})^{(m)}{\mathbfit{c}_{X,i}^{*}},\label{alphY}\\
&\boldsymbol{\Sigma}_{Y,k}^{(m)}=\frac{1}{n_{k}^{(m)}}\sum_{i=1}^n t_{ik}^{(m)}\biggl(wi_{ik,Y}^{(m)}\left({\mathbfit{c}}_{Y,i}-(\boldsymbol{\Gamma}^k_{*})^{(m)}{\mathbfit{c}_{X,i}^{*}} \right)\left({\mathbfit{c}}_{Y,i}-(\boldsymbol{\Gamma}^k_{*})^{(m)}{\mathbfit{c}_{X,i}^{*}} \right)^\top\notag\\
&+w_{ik,Y}^{(m)}\boldsymbol{\alpha}_{Y,k}\boldsymbol{\alpha}_{Y,k}^\top-\left(\mathbfit{c}_{Y,i}-\boldsymbol{\Gamma}^k_{*}{\mathbfit{c}_{X,i}^{*}}\right)\boldsymbol{\alpha}_{Y,k}^\top-\boldsymbol{\alpha}_{Y,k}\left(\mathbfit{c}_{Y,i}-\boldsymbol{\Gamma}^k_{*}{\mathbfit{c}_{X,i}^{*}}\right)^\top\biggl)\label{sigmY}
\end{align}
\begin{itemize}
\item If $\mathbfit{c}_{X,i}\mid Z_{ik}=1\sim VG_{R^X}(\boldsymbol{\mu}_{X,k},\boldsymbol{\alpha}_{X,k},\boldsymbol{\Sigma}_{X,k},\psi_{X,k})$, $k=1,\ldots, K$, then
the update $\psi_{X,k}^{(m)}$ is the solution of the equation
\begin{equation}
\log(\psi_{X,k})+1-DG(\psi_{X,k})+\bar{lw}_{k,X}^{(m)}-\bar{w}_{k,X}^{(m)}=0,\label{specparVGx1}
\end{equation}
where $DG(w)=\frac{d\log(\Gamma(x))}{dx}$ is the digamma function.
If $\psi_{X,k}=\psi_{X}$, $k=1,\ldots, K$, then $\psi_{X}$ is the solution of the equation
\begin{equation}
\log(\psi_{X})+1-DG(\psi_{X})+\frac{1}{n}\sum_{i=1}^n\sum_{k=1}^K t_{ik}^{(m)}\left( lw_{ik,X}^{(m)}-w_{ik,X}^{(m)}\right)=0.\label{specparVGx2}
\end{equation}
Similarly, if $\mathbfit{c}_{Y,i}\mid Z_{ik}=1, \mathbfit{c}_{X,i}\sim VG_{R^Y}(\boldsymbol{\mu}_{Y,k},\boldsymbol{\alpha}_{Y,k},\boldsymbol{\Sigma}_{Y,k},\psi_{Y,k})$, $k=1,\ldots, K$, then  $\psi_{Y,k}^{(m)}$ is the solution of the equation
\begin{equation}
\log(\psi_{Y,k})+1-DG(\psi_{Y,k})+\bar{lw}_{k,Y}^{(m)}-\bar{w}_{k,Y}^{(m)}=0,\label{specparVGy1}
\end{equation}
If $\psi_{Y,k}=\psi_Y$, $k=1,\ldots, K$, then $\psi_{Y}$ is the solution of the equation
\begin{equation}
\log(\psi_{Y})+1-DG(\psi_{Y})+\frac{1}{n}\sum_{i=1}^n\sum_{k=1}^K t_{ik}^{(m)}\left( lw_{ik,Y}^{(m)}-w_{ik,Y}^{(m)}\right)=0.\label{specparVGy2}
\end{equation}
\item If $\mathbfit{c}_{X,i}\mid Z_{ik}=1\sim ST_{R^X}(\boldsymbol{\mu}_{X,k},\boldsymbol{\alpha}_{X,k},\boldsymbol{\Sigma}_{X,k},\nu_{X,k})$, $k=1,\ldots, K$, then
 the update $\nu_{X,k}^{(m)}$ is the solution of the equation
\begin{equation}
\log\left(\frac{\nu_{X,k}}{2}\right)+1-DG\left(\frac{\nu_{X,k}}{2}\right)-\bar{lw}_{k,X}^{(m)}-\bar{wi}_{k,X}^{(m)}=0.\label{specparSTx1}
\end{equation}
If $\nu_{X,k}=\nu_{X}$, $k=1,\ldots, K$, then $\nu_{X}$ is the solution of the equation
\begin{equation}
\log\left(\frac{\nu_{X}}{2}\right)+1-DG\left(\frac{\nu_{X}}{2}\right)-\frac{1}{n}\sum_{i=1}^n\sum_{k=1}^K t_{ik}^{(m)}\left(lw_{ik,X}^{(m)}+wi_{ik,X}^{(m)}\right)=0.\label{specparSTx2}
\end{equation}
Similarly, if $\mathbfit{c}_{Y,i}\mid Z_{ik}=1, \mathbfit{c}_{X,i}\sim ST_{R^Y}(\boldsymbol{\mu}_{Y,k},\boldsymbol{\alpha}_{Y,k},\boldsymbol{\Sigma}_{Y,k},\nu_{Y,k})$, $k=1,\ldots, K$, then  $\nu_{Y,k}^{(m)}$ is the solution of the equation
\begin{equation}
\log\left(\frac{\nu_{Y,k}}{2}\right)+1-DG\left(\frac{\nu_{Y,k}}{2}\right)-\bar{lw}_{k,Y}^{(m)}-\bar{wi}_{k,Y}^{(m)}=0.\label{specparSTy1}
\end{equation}
If $\nu_{Y,k}=\nu_{Y}$, $k=1,\ldots, K$, then $\nu_{Y}$ is the solution of the equation
\begin{equation}
\log\left(\frac{\nu_{Y}}{2}\right)+1-DG\left(\frac{\nu_{Y}}{2}\right)-\frac{1}{n}\sum_{i=1}^n\sum_{k=1}^K t_{ik}^{(m)}\left(lw_{ik,Y}^{(m)}+wi_{ik,Y}^{(m)}\right)=0.\label{specparSTy2}
\end{equation}
\item If $\mathbfit{c}_{X,i}\mid Z_{ik}=1\sim NIG_{R^X}(\boldsymbol{\mu}_{X,k},\boldsymbol{\alpha}_{X,k},\boldsymbol{\Sigma}_{X,k},\kappa_{X,k})$, $k=1,\ldots, K$, then
\begin{equation}
\kappa_{X,k}^{(m)}=\frac{1}{\bar{w}_{k,X}^{(m)}}.\label{specparNIGx1}
\end{equation}
If $\kappa_{X,k}=\kappa_{X}$, $k=1,\ldots, K$, then  
\begin{equation}
\kappa_{X}=\frac{n}{\sum_{i=1}^n\sum_{k=1}^Kt_{ik}^{(m)}w_{ik,X}^{(m)}}.\label{specparNIGx2}
\end{equation}
Similarly, if $\mathbfit{c}_{Y,i}\mid Z_{ik}=1, \mathbfit{c}_{X,i}\sim NIG_{R^Y}(\boldsymbol{\mu}_{Y,k},\boldsymbol{\alpha}_{Y,k},\boldsymbol{\Sigma}_{Y,k},\kappa_{Y,k})$, $k=1,\ldots, K$, then 
\begin{equation}
\kappa_{Y,k}^{(m)}=\frac{1}{\bar{w}_{k,Y}^{(m)}}.\label{specparNIGy1}
\end{equation} 
If $\kappa_{Y,k}=\kappa_{Y}$, $k=1,\ldots, K$, then  
\begin{equation}
\kappa_{Y}=\frac{n}{\sum_{i=1}^n\sum_{k=1}^Kt_{ik}^{(m)}w_{ik,Y}^{(m)}}.\label{specparNIGy2}
\end{equation}
\end{itemize}
\end{proposition}
\begin{proof}
 The proof is included in  Appendix \ref{secA3}.
 \end{proof}
\subsubsection{Initialization and computational details}
We start the EM algorithm with initial guesses for some of the parameters. Depending on the pair of skewed distributions considered, we initialize the distribution specific parameters as $\phi_{X,k}^{(0)}=\nu_{X,k}^{(0)}=\kappa_{X,k}^{(0)}=\phi_{Y,k}^{(0)}=\nu_{Y,k}^{(0)}=\kappa_{Y,k}^{(0)}=10$, and all the entries of  $\boldsymbol{\alpha}_{X,k}^{(0)}$ and $\boldsymbol{\alpha}_{Y,k}^{(0)}$ are initialized with 10. We tried different values and the results were not influenced much by this initial guess.  On the other hand, the convergence of the EM algorithm is dependent on the initial values of  the parameters $t_{ik}^{(0)}$. We have implemented a random initialization and an  initialization with the {\it{kmeans}} method  (available in the {\it{stats}} package in R) applied to the data set formed by the combining  the coefficients  $\mathbfit{C_X}$, $\mathbfit{C_Y}$. To finalize the first E-step we use the {\it{cov.wt}} in the {\it{stats}} package in R to estimate $\mathbfit{S}_{X,k}^{(0)}$, $\boldsymbol{\Sigma}_{Y,k}^{(0)}$, $\boldsymbol{\mu}_{X,k}^{(0)}$, $\boldsymbol{\mu}_{Y,k}^{(0)}$ with the weights given by the initial values $t_{ik}^{(0)}$.
 
To prevent the convergence of the EM algorithm to a local maximum, we execute the algorithm multiple times, with different initialization values for $t_{ik}^{(0)}$. We keep the best result given by the EM algorithm using the Bayesian information criterion (BIC; \citealp{Schwarz:1978}) defined by 
\begin{equation}
BIC=L^{(m_f)}-\frac{\tau}{2}\log n,\label{bicm}
\end{equation}
where  $n$ is the number of observations, $\tau$ is the overall number of the free parameters, $L^{(m_f)}$ is the maximum log-likelihood value, and $m_f$ is the last iteration of the algorithm before convergence. The number of clusters $K$ and the parsimonious model are also selected by maximizing the the BIC.  As in \cite{DangPunzoMcNicholasIngrassiaBrowne:2017} we remove the models that have a matrix $\boldsymbol{\Sigma}_{Y,k}^{(m_f)}$ for which at least one eigenvalue  is less than $10^{-20}$ to disregard models with spurious clusters. 

We select the group specific dimension $d_k$ through the Cattell scree-test with a given threshold $\epsilon$ \citep{BouveyronJacques:2011}. The BIC can be used to select the optimum threshold, too. Alternatively to the Cattell scree-test, $d_k$  can be chosen through a grid search as the value corresponding to the maximum BIC \citep{Amovin-AssagbaGannazJacques:2022}. 

We determine the clusters using the maximum {\it{a posteriori}} (MAP) rule: an observation $(\mathbfit{c}_{Y,i},\mathbfit{c}_{X,i})$ is assigned to the cluster $k\in\{1,\ldots, K\}$ with the largest $t_{ik}^{(m_f)}$, where $m_f$ is the last iteration of the  EM algorithm before convergence. The EM algorithm is stopped after a maximum number of iterations, or when the difference $\mid L^{(m+2)}_{\infty}-L^{(m+1)}\mid<\epsilon_1$ \citep{McNicholasMurphyMcDaidFrost:2010}, where $L^{(m+1)}$ is the log-likelihood value at iteration $m+1$, and $L^{(m+2)}_{\infty}$ is the asymptotic estimate of log-likelihood  at iteration $m+2$ as defined in \cite{AndrewsMcNicholasSubedi:2011}. We choose 200 as maximum number of iterations, and the threshold $\epsilon_1=10^{-6}$. 




  
\section{Numerical Experiments}
\label{section4}
We apply the proposed clustering method to simulated data and to the Air Quality dataset  available in the {\it FRegSigCom} R package that contains the hourly averages of the concentration values of five atmospheric pollutants and the hourly average temperatures and humidity measured in a significantly polluted area, at road level, within an Italian city \citep{QiLuo:2019}.
\subsection{Simulated data}
We apply the proposed  method funWeightClusSkew to simulated data and we compare it with other methods for clustering functional data:  funWeightClust  \citep{AntonSmith:2025} that considers linear functional regression models based on multivariate normal distribution, and the methods  funHDDC and TfunHDDC that do not include covariates and are implemented  in the {\it funHDDC}  and {\it TfunHDDC} R packages, respectively. For the simulated data the true classifications are known, so we use the Adjusted Rand Index (ARI; \citealp{HubertArabie:1985}) to measure the accuracy of the classification. A perfect classification has ARI=1, the expected value of ARI is 0, and a classification with ARI$<0$ is worse than random assignment.

We consider four combinations, three of them with probability density functions $f_k(\mathbfit{c}_{X,i} \mid \boldsymbol{\theta}_{X,k})$ and $g_k(\mathbfit{c}_{Y,i} \mid \mathbfit{c}_{X,i},\boldsymbol{\theta}_{Y,k})$ of the same skewed type, and one with probability density functions of a different skewed type. We consider the following combinations: (1)NIG-VG, (2) NIG-NIG, (3) ST-ST, (4) VG-VG.  For each model we simulate 600 curves from 2 clusters with mixing proportions $\pi_1=\pi_2=1/2$. We simulate the coefficients in each case according to the corresponding skewed distributions and then curves are smoothed using 6 cubic B-spline basis functions. We run 100 simulations for each of the four cases.

\begin{figure}[h!]
\begin{center}
\includegraphics[width=14cm,height=5cm]{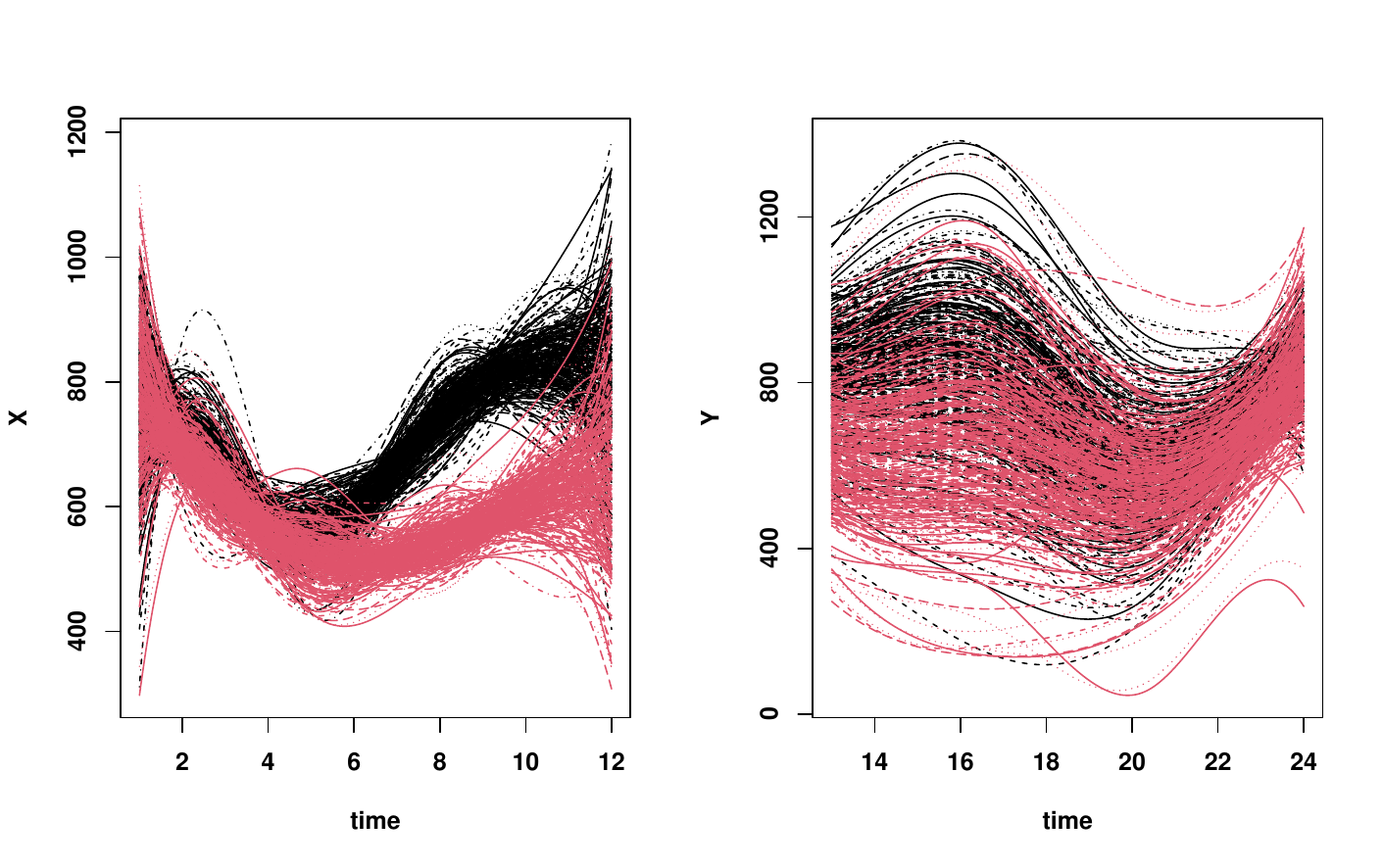}
\end{center}
  \caption{Smooth data  from NIG-VG distributions  colored by group for one simulation.}
\label{figNIGVG}
\end{figure}   
\begin{figure}[h!]
\begin{center}
\includegraphics[width=14cm,height=5cm]{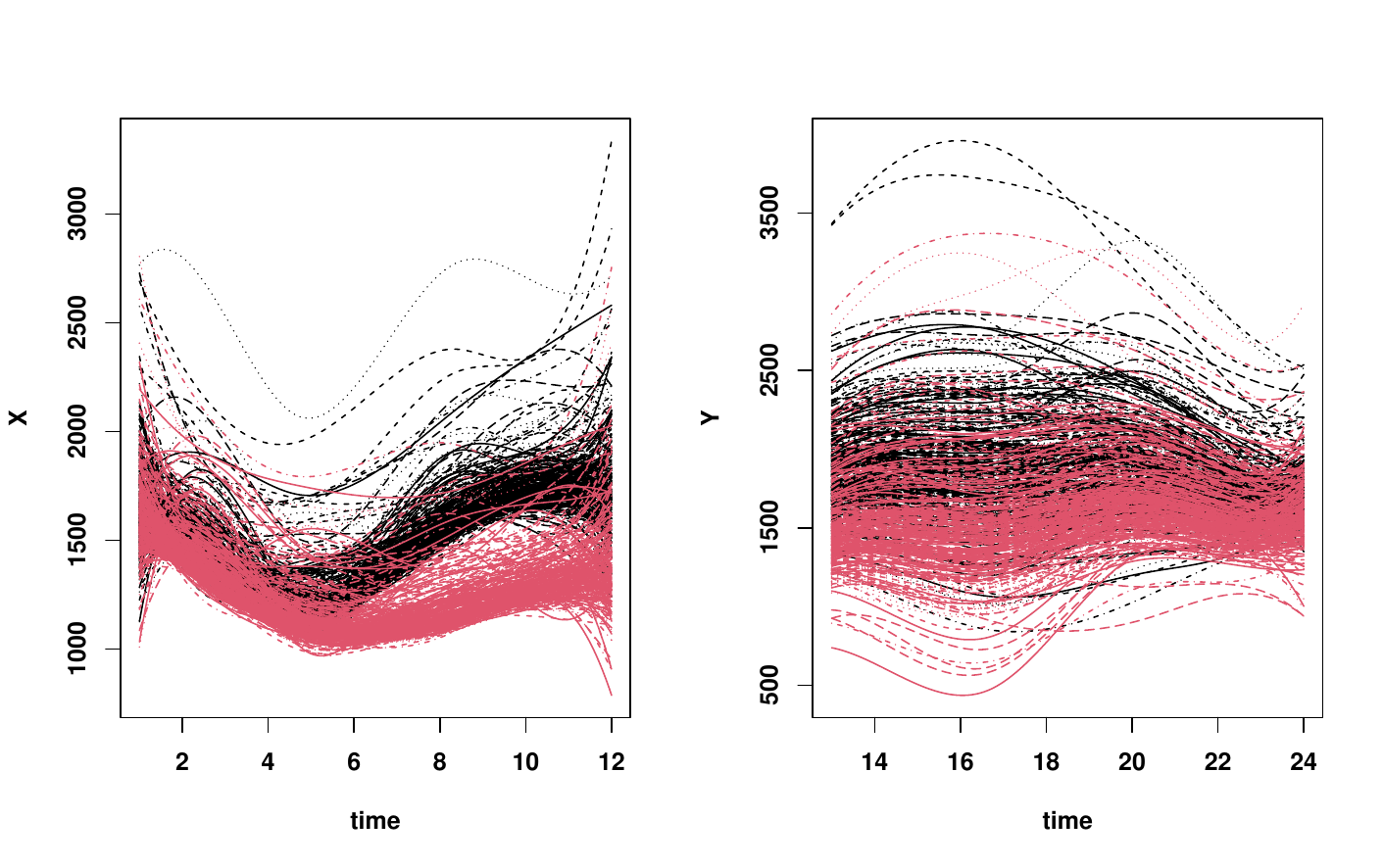}
\end{center}
  \caption{Smooth data  from NIG-NIG distributions  colored by group for one simulation.}
\label{figNIGNIG}
\end{figure} 
 \begin{figure}[h!]
\begin{center}
\includegraphics[width=14cm,height=5cm]{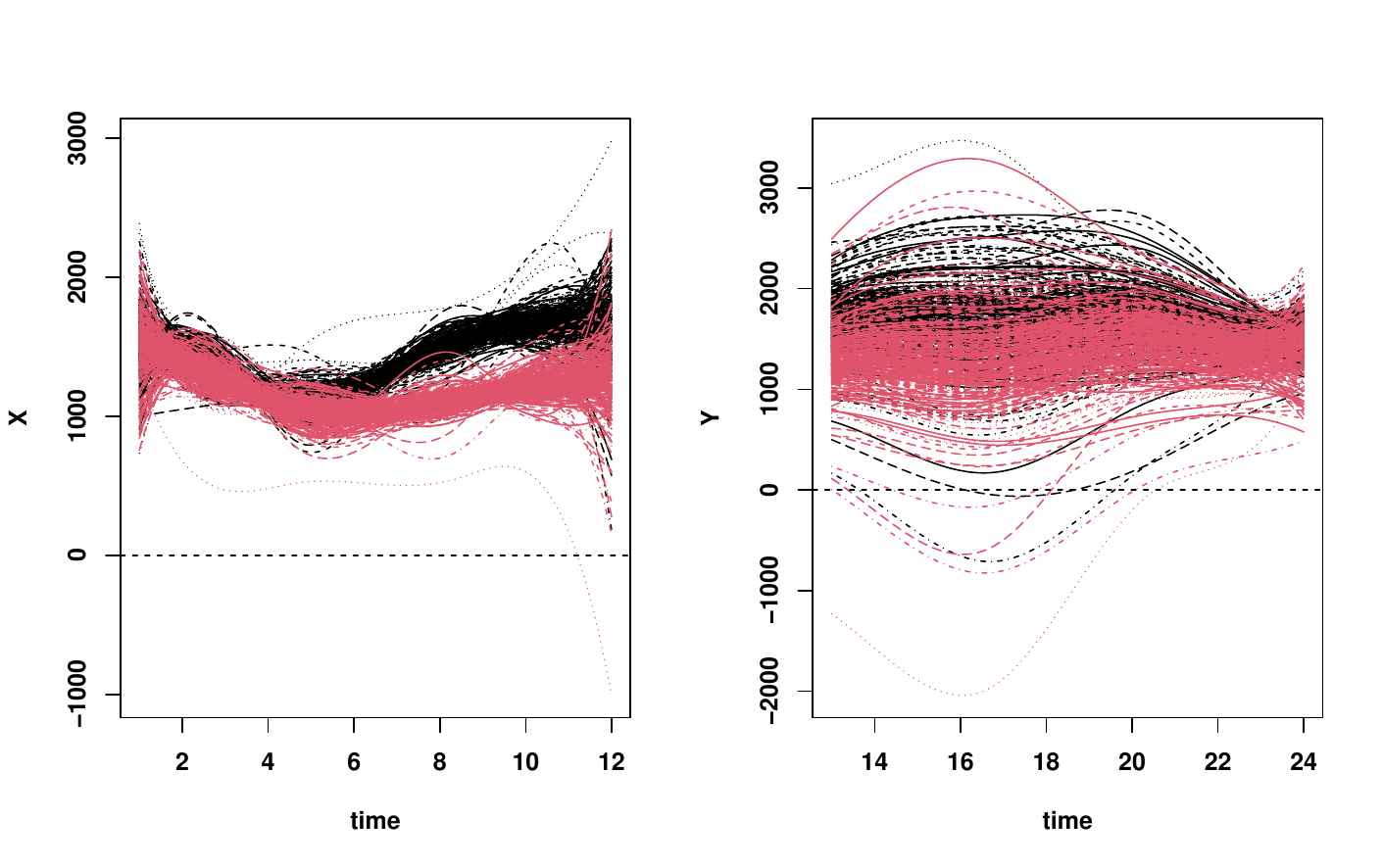}
\end{center}
  \caption{Smooth data from ST-ST distributions  colored by group for one simulation.}
\label{figSTST}
\end{figure}  
\begin{figure}[h!]
\begin{center}
\includegraphics[width=14cm,height=5cm]{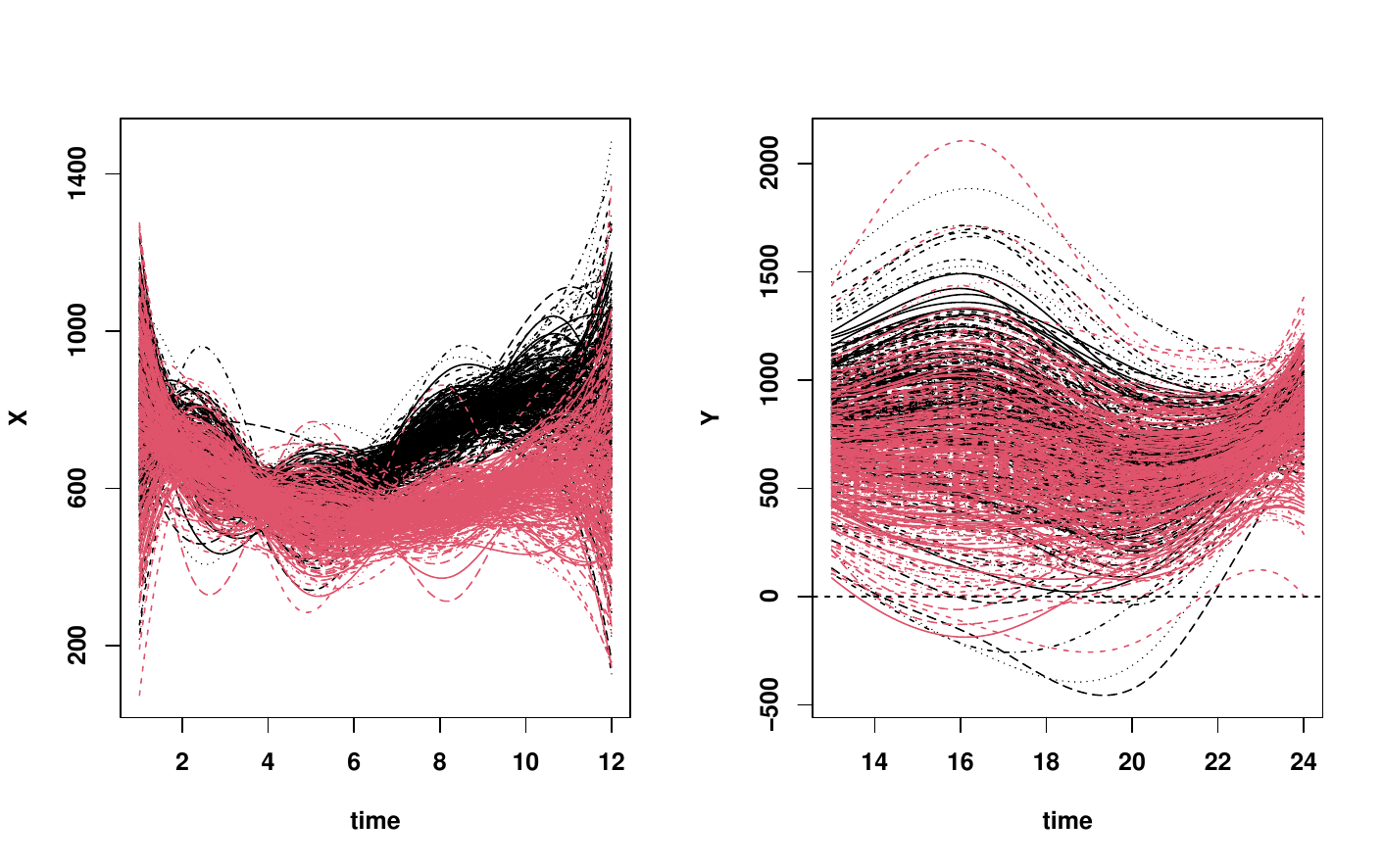}
\end{center}
  \caption{Smooth data simulated from VG-VG distributions  colored by group for one simulation.}
\label{figVGVG}
\end{figure}   
For the NIG-VG, NIG-NIG, and ST-ST scenarios we put $\boldsymbol{\Sigma}_{Y,1}=\boldsymbol{\Sigma}_{Y,2}=879.1197 \mathbf{I}_6$, where $\mathbf{I}_6$ is the six dimensional identity matrix. We  use the values for $\boldsymbol{\mu}_{X,1}$, $\boldsymbol{\mu}_{X,2}$,$\boldsymbol{\Sigma}_{X,1}$,  $\boldsymbol{\Sigma}_{X,2}$, $\boldsymbol{\Gamma}^1$, $\boldsymbol{\Gamma}^2$, $\boldsymbol{\Gamma}^1_0$, and $\boldsymbol{\Gamma}^2_0$, as given in Appendix \ref{apendD}.   For the NIG-VG case we have a NIG distribution with $\kappa_{X,1}=\kappa_{X,2}=3$ and $\boldsymbol{\alpha}_
{X,1}=\boldsymbol{\alpha}_{X,2}=0.1\mathbf{E}_6$, and a VG distribuion with  $\psi_{Y,1}=\psi_{Y,2}=2$ and $\boldsymbol{\alpha}_{Y,1}=\boldsymbol{\alpha}_{Y,2}=-0.5\mathbf{E}_6$, where $\mathbf{E}_6=(1,1,1,1,1,1)^\top$. In Figure \ref{figNIGVG} we plot the curves corresponding to one simulation. 
For the NIG-NIG case we consider NIG distributions with $\kappa_{X,1}=\kappa_{X,2}=1$ and $\kappa_{Y,1}=\kappa_{Y,2}=1.5$, $\boldsymbol{\alpha}_{X,1}=\boldsymbol{\alpha}_{X,2}=\boldsymbol{\alpha}_{Y,1}=\boldsymbol{\alpha}_{Y,2}=100\mathbf{E}_6$. One simulation is plotted in Figure \ref{figNIGNIG}. For the ST-ST case the parameters of the ST distributions are $\nu_{X,1}=\nu_{X,2}=4$ and $\nu_{Y,1}=\nu_{Y,2}=6$, $\boldsymbol{\alpha}_{X,1}=\boldsymbol{\alpha}_{X,2}=0.5\mathbf{E}_6$  and $\boldsymbol{\alpha}_{Y,1}=\boldsymbol{\alpha}_{Y,2}=-0.5\mathbf{E}_6$. In Figure \ref{figSTST} we plot one simulation.

For the VG-VG simulations we use a more complicated covariance structure with non-diagonal matrices $\boldsymbol{\Sigma}_{Y,1}=\boldsymbol{\Sigma}_{Y,1}$ given in Appendix \ref{apendD} and
\begin{align*}
&\boldsymbol{\alpha}_{X,1}=(0.5,-0.10, 0.25,-0.2,-0.5,1)^\top,\quad
\boldsymbol{\alpha}_{X,2}=(-0.5,2.50, 1.3,-1.50,0.5,1)^\top,\\
&\boldsymbol{\alpha}_{Y,1}=(-0.5,1.00,-1.50,0.20,0.5-1.5)^\top,\quad
\boldsymbol{\alpha}_{Y,2}=(0.5,-1.50,-0.1,0.3,0.1,-0.6)^\top.
\end{align*}
The values for $\boldsymbol{\mu}_{X,1}$, $\boldsymbol{\mu}_{X,2}$,$\boldsymbol{\Sigma}_{X,1}$,  $\boldsymbol{\Sigma}_{X,2}$, $\boldsymbol{\Gamma}^1$, $\boldsymbol{\Gamma}^2$, $\boldsymbol{\Gamma}^1_0$, and $\boldsymbol{\Gamma}^2_0$ are given in Appendix \ref{apendD}.
We consider VG distributions with $\psi_{X,1}=\psi_{X,2}=3$, $\psi_{Y,1}=\psi_{Y,2}=2$. In Figure \ref{figVGVG} we plot one simulation for the VG-VG distributions. 

Comparing the four scenarios in Figures \ref{figNIGVG}-\ref{figVGVG} we notice that for all of them there is a lot of ovarlapping for the response curves $Y_i$. For the predictor curves $X_i$, the least ovarlapping is for the NIG-VG distributions and the largest for the VG-VG distributions. 

We apply funWeightClustSkew, funWeightClust , funHDDC  and tfunHDDC and we use ARI to evaluate the performance. When we apply funHDDC  and tfunHDDC we consider the pairs of curves $(X_i,Y_i)$ as two-dimensional functional data.  We run all four methods for $K=2$, we use {{\it{kmeans}}} method  with 20 repetitions for initialization, and the maximum number of iterations is 200 for the stopping criterion. We optimize based on BIC for the threshold $\epsilon$ in the Cattell test,   For each method we  consider all sub-models and the best model is chosen as the one with the highest BIC value.
   
From the results included in Table \ref{table1} we notice that funWeightClustSkew outperforms the other methods. For scenario NIG-VG both tfunHDDC and funWeightClust also give good results. For scenario NIG-NIG, the methods funWeightClustSkew  and tfunHDDC based on non-normal distributions give the best results. For scenario ST-ST, taking into account the relationship between $Y_i$ and $X_i$ seems to matter and funWeightClust and funWeightClustSkew perform better than funHDDC  and tfunHDDC. The scenario VG-VG has the most overlapping and only funWeightClustSkew gives good results.

\begin{longtable}{@{}lllll@{}}
\caption{Mean (and standard deviation) of ARI for BIC best model on 100 simulations.}\\
\label{table1}
Scenario&Method &ARI Mean (St. Dev)&ARI -Median \\
\midrule
NIG-VG&FunHDDC& {\bf{0.19 (0.07)}}&{\bf{0.20}}\\
NIG-VG&tfunHDDC&{\bf{0.73 (0.12)}}&{\bf{0.72}}\\
NIG-VG&funWeightClust& {\bf{0.75 (0.3)}}&{\bf{0.95}}\\
NIG-VG&funWeightClustSkew& {\bf{0.92 (0.2)}}&{\bf{0.99}}\\
\midrule
NIG-NIG&FunHDDC& {\bf{0.31(0.1)}}&{\bf{0.33}}\\
NIG-NIG&tfunHDDC&{\bf{0.78(0.09)}}&0.80\\
NIG-NIG&funWeightClust& {\bf{0.54 (0.22)}}&{\bf{0.50}}\\
NIG-NIG&funWeightClustSkew& {\bf{0.85(0.06)}}&{\bf{0.86}}\\
\midrule
ST-ST&FunHDDC& {\bf{0.26 (0.06)}}&{\bf{0.27}}\\
ST-ST&tfunHDDC&{\bf{0.61 (0.14)}}&{\bf{0.6}}\\
ST-ST&funWeightClust& {\bf{0.70 (0.17)}}&{\bf{0.94}}\\
ST-ST&funWeightClustSkew& {\bf{0.88(0.10)}}&0.90\\
\midrule
VG-VG&FunHDDC& {\bf{0.10 (0.03)}}&{\bf{0.10}}\\
VG-VG&tfunHDDC&{\bf{0.14 (0.07)}}&{\bf{0.14}}\\
VG-VG&funWeightClust& {\bf{0.06 (0.1)}}&0.01\\
VG-VG&funWeightClustSkew& {\bf{0.95(0.12)}}&{\bf{0.99}}\\
\midrule
\end{longtable}
\subsection{The daily air quality dataset}
We consider the daily air quality data set available in the R package {\it FRegSigCom} and consisting of 355 daily curves for each of the  pollutants:  nitrogen dioxide
(NO$_2$), carbon monoxide (CO), non-methane hydrocarbons (NMHC), total
nitrogen oxides (NO$_x$), and benzene (C$_6$H$_6$) \citep{DeVito:2008}. The measurements were done in a polluted area of an Italian city polluted area, at road level, and the curves represent the hourly averages of the concentration values for the five pollutants. In \cite{QiLuo:2019} assuming a nonlinear relationship, the daily curve of NO$_2$ is predicted by the daily curves of the other four pollutants together with the temperature and relative humidity. Since a change of the relationship between NO$_2$ and NO$_x$ is observed depending on different  concentration levels of NO$_x$, here we cluster the curves for these five pollutants based on the relationship between the concentrations of NO$_2$ and the concentrations of the other four pollutants,  temperature and relative humidity. We use a B-spline basis with 12 splines.
\begin{figure}[h!]
\begin{center}
\includegraphics[width=14cm,height=9cm]{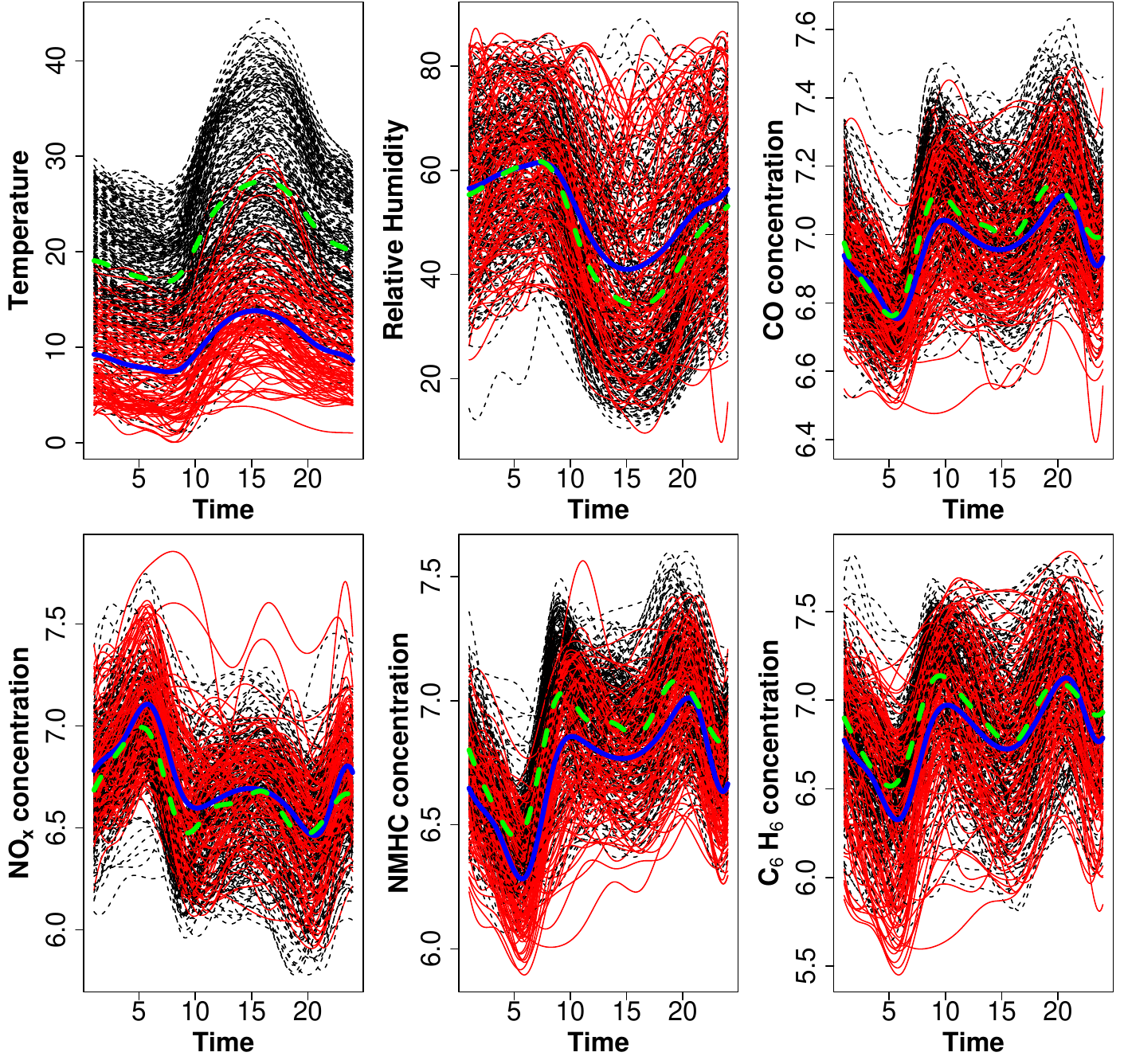}
\end{center}
  \caption{The 355 daily curves for each of the six predictors in the Air Quality dataset colored by group and the group estimated means.}
\label{figpoluantsp1}
\end{figure}   
\begin{figure}[h!]
\begin{center}
\includegraphics[width=14cm,height=5cm]{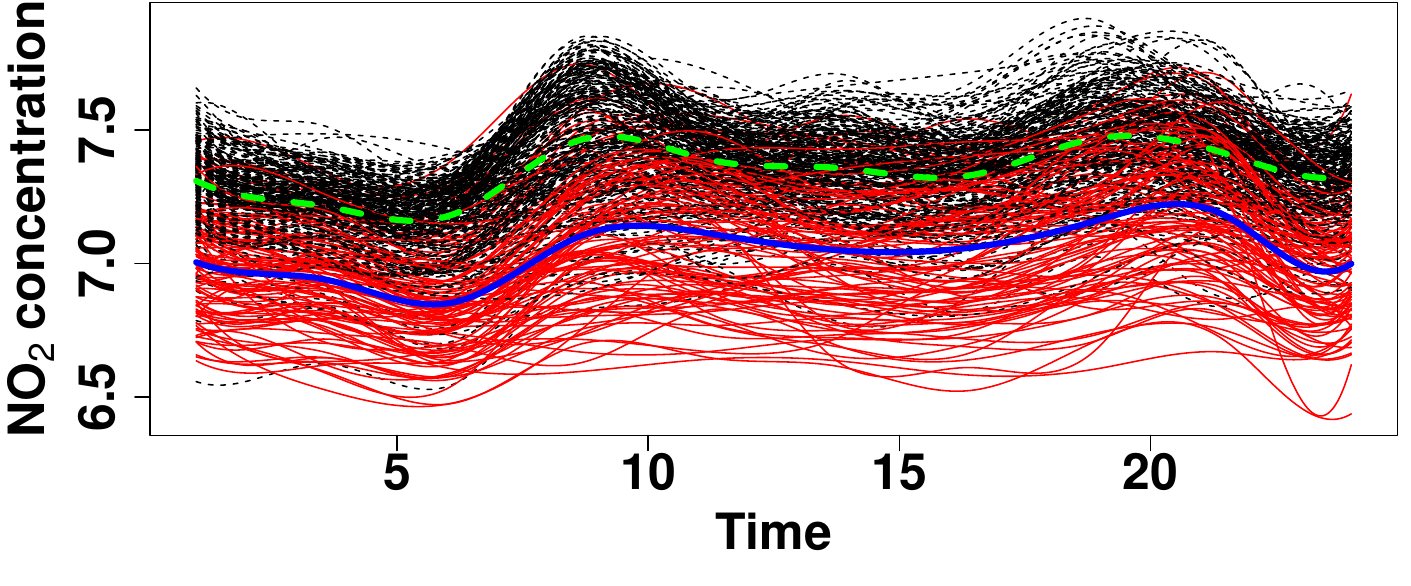}
\end{center}
  \caption{The 355 daily curves of NO$_2$ in the Air Quality dataset colored by group and the group estimated means.}
\label{figpoluantsr1}
\end{figure}   

We apply the method funWeightClustSkew  with the number of clusters $K=2,\ldots, 6$ and based on the BIC we choose a model with two clusters. The distribution of the predictors is VG with the parsimonious model  ABQKDK and the distribution for the response variable is NIG with the parsimonious model EVI. We choose the best value based on the BIC for the threshold  for the scree-test of Cattell. In Figure \ref{figpoluantsp1} we plot the curves for the six predictors and in Figure \ref{figpoluantsr1} we plot daily curves of NO$_2$. The curves for the first cluster are plotted with black dashed lines and the curves for the second cluster with red plain lines. The estimated mean curve for the first cluster is plotted with a thick dashed green line and for the second cluster we use a thick plain blue line.

From Figure \ref{figpoluantsr1}  we notice that the first cluster includes mainly the curves corresponding to higher concentrations of  the response variable NO$_2$. For the predictors curves, this cluster corresponds to warmer days (see the top left picture in Figure \ref{figpoluantsp1}). There is a visible difference in terms of temperature between the two clusters, with the average temperature for the first cluster at 21.72C and for the second cluster at 10.23C. For the other five predictors, there is a lot of overlapping and there is no obvious interpretation of the clustering.

 
\section{Conclusions and future work}
\label{section5}  
We propose the method funWeightClustSkew for clustering heterogeneous functional linear regression data. This extends  funWeightClust \citep{AntonSmith:2025} by including, in addition to the multivariate normal distribution,  three multivariate skewed distributions: the variance-gamma distribution , the  skew-t  distribution, and the normal-inverse Gaussian distribution.  It is also a functional data version of the method proposed in \cite{GallaugherTomarchioMcNicholasPunzo:2022} for multivariate data. As mentioned in \cite{ BouveyronJacques:2011}, applying a multivariate data method to a discretization of the functional data has the disadvantage that the results depend on the chosen discretization, so by considering the functional nature of the data funWeightClustSkew is a good alternative approach. 

We conduct experiments for simulated data and compare the proposed method with funWeightClust,  funHDDC  and tfunHDDC.  Since these simulated data include linear regression dependencies between the variables and non-normal distributions, funHDDC has the worst performance. FunWeightClust  and tfunHDDC have similar performances, but they are outperformed by  funWeightClustSkew. We also use funWeightClustSkew to cluster the daily Air Quality dataset \citep{DeVito:2008}. From the clustering we can see that the low and high levels of the concentration of NO$_2$  give different kinds of dependencies  on temperature, relative humidity, and the concentrations of four other pollutants. 

The proposed model can be also applied for functional classification and prediction (\citealp{ChiouYangChen:2016}, \citealp{Chiou:2012}). Moreover, the EM algorithm can be modified to include clustering with missing data \citep{TongTartora:2022}. 


\begin{appendices}
\section{Proposition \ref{proplik}}\label{secA1}
\begin{proposition}
 \label{proplik}
 The complete data log-likelihood of the observed curves under the FLM[$a_{kj},b_k,\mathbfit{Q}_k,d_k$] - VVV model   can be written as
 \begin{equation}
l_c(\boldsymbol{\theta})=l_{1c}(\pi)+l_{2c}(\boldsymbol{\vartheta}_X)+l_{3c}(\boldsymbol{\vartheta}_Y)+l_{4c}(\theta_{W,X}\cup \theta_{W,Y})\label{lik2}
\end{equation}
where
\begin{align}
&l_{1c}(\pi)=\sum_{i=1}^n\sum_{k=1}^K z_{ik}\log(\pi_k),\label{l1c2}\\
&l_{2c}(\boldsymbol{\vartheta}_X)=-\frac{1}{2}\sum_{i=1}^n\log(w_{i,X})-\frac{nR_X\log(2\pi)}{2}\notag\\
&+\frac{n}{2}\log(\mid \mathbfit{W}_X\mid)-\frac{1}{2}\sum_{k=1}^K n_k\sum_{l=1}^{d_k}\log(a_{kl})-\frac{1}{2}\sum_{k=1}^K n_k\sum_{l=d_k+1}^{R_X}\log(b_{k})\notag\\
&-\frac{1}{2}\sum_{k=1}^K   \biggl(\sum_{l=1}^{d_k}\frac{\mathbfit{q}_{kl}^\top \mathbfit{W}_X^{1/2}\mathbfit{S}_{X,k}\mathbfit{W}_X^{1/2}\mathbfit{q}_{kl}}{a_{kl}}+\sum_{l=d_k+1}^{R_X}\frac{\mathbfit{q}_{kl}^\top \mathbfit{W}_X^{1/2}\mathbfit{S}_{X,k}\mathbfit{W}_X^{1/2}\mathbfit{q}_{kl}}{b_{k}}\biggl)\label{l3c2}\\
&l_{3c}(\boldsymbol{\vartheta}_Y)=-\frac{1}{2}\sum_{i=1}^n\log(w_{i,Y})-\frac{nR_Y\log(2\pi)}{2}\notag\\
&-\frac{1}{2}\sum_{k=1}^K n_k\log(\log\mid \boldsymbol{\Sigma}_{Y,k}\mid )-\frac{1}{2}\sum_{i=1}^n\sum_{k=1}^K \frac{z_{ik}}{w_{i,Y}}\biggl(\mathbfit{c}_{Y,i}^\top\boldsymbol{\Sigma}_{Y,k}^{-1}\mathbfit{c}_{Y,i}-\mathbfit{c}_{Y,i}^\top\boldsymbol{\Sigma}_{Y,k}^{-1}\boldsymbol{\Gamma}^k_{*}{\mathbfit{c}_{X,i}^{*}}\notag\\
&-({\mathbfit{c}_{X,i}^{*}})^\top\left(\boldsymbol{\Gamma}^k_{*}\right)^\top \boldsymbol{\Sigma}_{Y,k}^{-1}\mathbfit{c}_{Y,i}+({\mathbfit{c}_{X,i}^{*}})^\top\left(\boldsymbol{\Gamma}^k_{*}\right)^\top\boldsymbol{\Sigma}_{Y,k}^{-1}\boldsymbol{\Gamma}^k_{*}{\mathbfit{c}_{X,i}^{*}}\biggl)-\frac{1}{2}\sum_{i=1}^n\sum_{k=1}^K z_{ik}w_{i,Y}\boldsymbol{\alpha}_{Y,k}^\top\boldsymbol{\Sigma}_{Y,k}^{-1} \boldsymbol{\alpha}_{Y,k}\notag\\
&+\frac{1}{2}\sum_{i=1}^n\sum_{k=1}^K z_{ik}\biggl(\mathbfit{c}_{Y,i}^\top\boldsymbol{\Sigma}_{Y,k}^{-1}\boldsymbol{\alpha}_{Y,k}-\left({\mathbfit{c}_{X,i}^{*}}\right)^\top\left(\boldsymbol{\Gamma}^k_{*}\right)^\top\boldsymbol{\Sigma}_{Y,k}^{-1}\boldsymbol{\alpha}_{Y,k}+\boldsymbol{\alpha}_{Y,k}^\top\boldsymbol{\Sigma}_{Y,k}^{-1}\mathbfit{c}_{Y,i}-\boldsymbol{\alpha}_{Y,k}^\top\boldsymbol{\Sigma}_{Y,k}^{-1}\boldsymbol{\Gamma}^k_{*}{\mathbfit{c}_{X,i}^{*}}\biggl)\label{l2c2}\\
&l_{4c}(\theta_{W,X}\cup \theta_{W,Y})=\sum_{i=1}^n\sum_{k=1}^K z_{ik}\left(\log (h(w_{i,X};\theta_{W,X}))+\log (h(w_{i,Y};\theta_{W,Y}))\right)\label{l4c2}
\end{align}
where  $\boldsymbol{\vartheta}_X=\{\boldsymbol{\mu}_{X,k}, a_{kj}, b_k, \mathbfit{q}_{kj},\boldsymbol{\alpha}_{X,k},\theta_{W,X}\}$, $\boldsymbol{\vartheta}_Y=\{\boldsymbol{\Sigma}_{Y,k}, \boldsymbol{\Gamma}^k_{*}, \boldsymbol{\alpha}_{Y,k},\theta_{W,Y} \}$, $k=1,\ldots, K$, $j=1,\ldots, d_k$, with $\mathbfit{q}_{kj}$ the $j$th column of $\mathbfit{Q}_k$, and $\theta_{W,X}$, $\theta_{W,Y}$ contain the parameters $\psi$, $\nu$, $\kappa$ specific to the distribution considered. Here the densities $h$ are given in formulas \eqref{h1l}, \eqref{h2l}, or \eqref{h3l}, according to the distribution considered,  $n_k=\sum_{i=1}^n z_{ik}$, and $\mathbfit{S}_{X,k}$ is defined by
\begin{equation}
\mathbfit{S}_{X,k}:=\sum_{i=1}^n \frac{z_{ik}}{w_{i,X}}(\mathbfit{c}_{X,i}-\boldsymbol{\mu}_{X,k}-w_{i,X}\boldsymbol{\alpha}_{X,k})(\mathbfit{c}_{X,i}-\boldsymbol{\mu}_{X,k}-w_{i,X}\boldsymbol{\alpha}_{X,k})^\top. \label{defS}
\end{equation}
 \end{proposition}
\begin{proof}
The complete-data likelihood can be written as the product of the conditional densities of the multivariate response $\mathbfit{c}_{Y,i}$ given the covariates ${\mathbfit{c}}_{X,i}$ and $\mathbfit{Z}_i=\mathbfit{z}_i$, $W_{Y,i}=w_{i,Y}$, the conditional densities of $W_{Y,i}$ given that $\mathbfit{Z}_i=\mathbfit{z}_i$, the conditional densities of ${\mathbfit{c}}_{X,i}$ given that $\mathbfit{Z}_i=\mathbfit{z}_i$, $W_{X,i}=w_{i,x}$, the conditional densities of $W_{Y,i}$ given that $\mathbfit{Z}_i=\mathbfit{z}_i$, and the marginal densities of the  $\mathbfit{Z}_i$:
\begin{align*}
L_c(\boldsymbol{\theta})&=\prod_{i=1}^n\prod_{k=1}^K\biggl\{\phi({\mathbfit{c}}_{Y,i};\boldsymbol{\mu}_{Y,k}+w_{i,Y}\boldsymbol{\alpha}_{Y,k},w_{i,Y}\boldsymbol{\Sigma}_{Y,k})h(w_{i,Y};\theta_{W,Y})\\
&\phi({\mathbfit{c}}_{X,i};\boldsymbol{\mu}_{X,k}+w_{i,X}\boldsymbol{\alpha}_{X,k},w_{i,X}\boldsymbol{\Sigma}_{X,k})h(w_{i,X};\theta_{W,X})\pi_k\biggl\}^{z_{ik}},
\end{align*}
where $z_{ik}=1$ if $(\mathbfit{c}_{Y,i}, \mathbfit{c}_{X,i})$ belongs to the cluster $k$ and $z_{ik}=0$ otherwise.
Thus, the complete-data  log-likelihood can be written as
\begin{equation*}
l_c(\boldsymbol{\theta})=l_{1c}(\pi)+l_{2c}(\boldsymbol{\vartheta}_X)+l_{3c}(\boldsymbol{\vartheta}_Y)+l_{4c}(\theta_{W,X}\cup \theta_{W,Y})
\end{equation*}
where
\begin{align}
&l_{1c}(\pi)=\sum_{i=1}^n\sum_{k=1}^K z_{ik}\log(\pi_k)\nonumber\\
&l_{2c}(\boldsymbol{\vartheta}_X)=-\frac{1}{2}\sum_{i=1}^n\sum_{k=1}^K z_{ik}\biggl(R_X\log(2\pi)+\log\mid \boldsymbol{\Sigma}_{X,k}\mid +\log(w_{i,X})\notag\\
&+\frac{1}{w_{i,X}}(\mathbfit{c}_{X,i}-\boldsymbol{\mu}_{X,k})^\top\boldsymbol{\Sigma}_{X,k}^{-1}(\mathbfit{c}_{X,i}-\boldsymbol{\mu}_{X,k})+w_{i,X}\boldsymbol{\alpha}_{X,k}^\top\boldsymbol{\Sigma}_{X,k}^{-1} \boldsymbol{\alpha}_{X,k}-(\mathbfit{c}_{X,i}-\boldsymbol{\mu}_{X,k})^\top\boldsymbol{\Sigma}_{X,k}^{-1}\boldsymbol{\alpha}_{X,k}\notag\\
&-\boldsymbol{\alpha}_{X,k}^\top\boldsymbol{\Sigma}_{X,k}^{-1}(\mathbfit{c}_{X,i}-\boldsymbol{\mu}_{X,k})\biggl)\label{l3use}\\
&l_{3c}(\boldsymbol{\vartheta}_Y)=-\frac{1}{2}\sum_{i=1}^n\sum_{k=1}^K z_{ik}\biggl(R_Y\log(2\pi)+\log\mid \boldsymbol{\Sigma}_{Y,k}\mid+ \log(w_{i,Y})\notag\\
&+\frac{1}{w_{i,Y}}(\mathbfit{c}_{Y,i}-\boldsymbol{\mu}_{Y,k})^\top\boldsymbol{\Sigma}_{Y,k}^{-1}(\mathbfit{c}_{Y,i}-\boldsymbol{\mu}_{Y,k})+w_{i,Y}\boldsymbol{\alpha}_{Y,k}^\top\boldsymbol{\Sigma}_{Y,k}^{-1} \boldsymbol{\alpha}_{Y,k}-(\mathbfit{c}_{Y,i}-\boldsymbol{\mu}_{Y,k})^\top\boldsymbol{\Sigma}_{Y,k}^{-1}\boldsymbol{\alpha}_{Y,k}\notag\\
&-\boldsymbol{\alpha}_{Y,k}^\top\boldsymbol{\Sigma}_{Y,k}^{-1}(\mathbfit{c}_{Y,i}-\boldsymbol{\mu}_{Y,k})\biggl)\nonumber\\
&=-\frac{1}{2}\sum_{i=1}^n\sum_{k=1}^K z_{ik}\biggl(R_Y\log(2\pi)+\log\mid \boldsymbol{\Sigma}_{Y,k}\mid+ \log(w_{i,Y})\notag\\
&+\frac{1}{w_{i,Y}}\left(\mathbfit{c}_{Y,i}-\boldsymbol{\Gamma}^k_{*}{\mathbfit{c}_{X,i}^{*}}\right)^\top\boldsymbol{\Sigma}_{Y,k}^{-1}\left(\mathbfit{c}_{Y,i}-\boldsymbol{\Gamma}^k_{*}{\mathbfit{c}_{X,i}^{*}}\right)+w_{i,Y}\boldsymbol{\alpha}_{Y,k}^\top\boldsymbol{\Sigma}_{Y,k}^{-1} \boldsymbol{\alpha}_{Y,k}\notag\\
&-\left(\mathbfit{c}_{Y,i}
-\boldsymbol{\Gamma}^k_{*}{\mathbfit{c}_{X,i}^{*}}\right)^\top\boldsymbol{\Sigma}_{Y,k}^{-1}\boldsymbol{\alpha}_{Y,k}
-\boldsymbol{\alpha}_{Y,k}^\top\boldsymbol{\Sigma}_{Y,k}^{-1}\left(\mathbfit{c}_{Y,i}-\boldsymbol{\Gamma}^k_{*}{\mathbfit{c}_{X,i}^{*}}\right)\biggl)\label{l2use}\\
&l_{4c}(\theta_{W,X}\cup \theta_{W,Y})=\sum_{i=1}^n\sum_{k=1}^K z_{ik}\left(\log (h(w_{i,X};\theta_{W,X}))+\log (h(w_{i,Y};\theta_{W,Y}))\right)\label{l4use}
\end{align}
From \eqref{dis5} we have 
$$
\boldsymbol{\Sigma}_{X,k}^{-1}=\mathbfit{W}_X^{1/2}\mathbfit{Q}_k\mathbfit{D}_k^{-1}\mathbfit{Q}_k^\top \mathbfit{W}_X^{1/2},
$$
and
\begin{equation} 
\mid \boldsymbol{\Sigma}_{X,k}\mid =\mid \mathbfit{D}_k\mid \mid \mathbfit{W}_X\mid ^{-1}\mid \mid \mathbfit{Q}_k^\top\mathbfit{Q}_k\mid  =\mid \mathbfit{D}_k\mid \mid \mathbfit{W}_X\mid ^{-1} =\mid \mathbfit{W}_X\mid ^{-1}\prod_{l=1}^{d_k}a_{kl}\prod_{l=d_k+1}^{R_X} b_k.\label{dety}
\end{equation}
Moreover, since $(\mathbfit{c}_{X,i}-\boldsymbol{\mu}_{X,k}-w_{i,X}\boldsymbol{\alpha}_{X,k})^\top\boldsymbol{\Sigma}_{X,k}^{-1}(\mathbfit{c}_{X,i}-\boldsymbol{\mu}_{X,k}-w_{i,X}\boldsymbol{\alpha}_{X,k})$ is a scalar, we get
\begin{align}
&(\mathbfit{c}_{X,i}-\boldsymbol{\mu}_{X,k}-w_{i,X}\boldsymbol{\alpha}_{X,k})^\top\boldsymbol{\Sigma}_{X,k}^{-1}(\mathbfit{c}_{X,i}-\boldsymbol{\mu}_{X,k}-w_{i,X}\boldsymbol{\alpha}_{X,k})\notag\\
&=\text{trace}\left((\mathbfit{c}_{X,i}-\boldsymbol{\mu}_{X,k}-w_{i,X}\boldsymbol{\alpha}_{X,k})^\top\mathbfit{W}_X^{1/2}\mathbfit{Q}_k\mathbfit{D}_k^{-1}\mathbfit{Q}_k^\top \mathbfit{W}_X^{1/2}(\mathbfit{c}_{X,i}-\boldsymbol{\mu}_{X,k}-w_{i,X}\boldsymbol{\alpha}_{X,k})\right)\notag
\\
&=\text{trace}\left(\left((\mathbfit{c}_{X,i}-\boldsymbol{\mu}_{X,k}-w_{i,X}\boldsymbol{\alpha}_{X,k})^\top\mathbfit{W}_X^{1/2}\mathbfit{Q}_k\right)\left(\mathbfit{D}_k^{-1}\mathbfit{Q}_k^\top \mathbfit{W}_X^{1/2}(\mathbfit{c}_{X,i}-\boldsymbol{\mu}_{X,k}-w_{i,X}\boldsymbol{\alpha}_{X,k})\right)\right)\nonumber\\
&=\text{trace}\left(\left(\mathbfit{D}_k^{-1}\mathbfit{Q}_k^\top \mathbfit{W}_X^{1/2}(\mathbfit{c}_{X,i}-\boldsymbol{\mu}_{X,k}-w_{i,X}\boldsymbol{\alpha}_{X,k})\right)\left((\mathbfit{c}_{X,i}-\boldsymbol{\mu}_{X,k}-w_{i,X}\boldsymbol{\alpha}_{X,k})^\top\mathbfit{W}_X^{1/2}\mathbfit{Q}_k\right)\right)\label{delty}
\end{align}
Replacing  in \eqref{l3use} we obtain
\begin{align*}
&l_{2c}(\boldsymbol{\vartheta}_X)=-\frac{1}{2}\sum_{i=1}^n\log(w_{i,X})-\frac{nR_X\log(2\pi)}{2}+\frac{n}{2}\log(\mid \mathbfit{W}_X\mid)\\
&-\frac{1}{2}\sum_{k=1}^K n_k\sum_{l=1}^{d_k}\log(a_{kl})-\frac{1}{2}\sum_{k=1}^K n_k\sum_{l=d_k+1}^{R_X}\log(b_{k})\\
&-\frac{1}{2}\sum_{k=1}^K   \biggl(\sum_{l=1}^{d_k}\frac{\mathbfit{q}_{kl}^\top \mathbfit{W}_X^{1/2}\mathbfit{S}_{X,k}\mathbfit{W}^{1/2}\mathbfit{q}_{kl}}{a_{kl}}+\sum_{l=d_k+1}^{R_X}\frac{\mathbfit{q}_{kl}^\top \mathbfit{W}_X^{1/2}\mathbfit{S}_{X,k}\mathbfit{W}_X^{1/2}\mathbfit{q}_{kl}}{b_{k}}\biggl),
\end{align*}
where $\mathbfit{q}_{kl}$ is the $l$th column of $\mathbfit{Q}_k$, and $\mathbfit{S}_{X,k}$ is defined in \eqref{defS}. 
Next, from \eqref{l2use} we have
\begin{align*}
&l_{3c}(\boldsymbol{\vartheta}_Y)=-\frac{1}{2}\sum_{i=1}^n\log(w_{i,Y})-\frac{nR_Y\log(2\pi)}{2}-\frac{1}{2}\sum_{k=1}^K n_k\log(\mid \boldsymbol{\Sigma}_{Y,k}\mid )\\
&-\frac{1}{2}\sum_{i=1}^n\sum_{k=1}^K \frac{z_{ik}}{w_{i,Y}}\biggl(\mathbfit{c}_{Y,i}^\top\boldsymbol{\Sigma}_{Y,k}^{-1}\mathbfit{c}_{Y,i}-\mathbfit{c}_{Y,i}^\top\boldsymbol{\Sigma}_{Y,k}^{-1}\boldsymbol{\Gamma}^k_{*}{\mathbfit{c}_{X,i}^{*}}-({\mathbfit{c}_{X,i}^{*}})^\top\left(\boldsymbol{\Gamma}^k_{*}\right)^\top \boldsymbol{\Sigma}_{Y,k}^{-1}\mathbfit{c}_{Y,i}\\
&+({\mathbfit{c}_{X,i}^{*}})^\top\left(\boldsymbol{\Gamma}^k_{*}\right)^\top\boldsymbol{\Sigma}_{Y,k}^{-1}\boldsymbol{\Gamma}^k_{*}{\mathbfit{c}_{X,i}^{*}}\biggl)-\frac{1}{2}\sum_{i=1}^n\sum_{k=1}^K z_{ik}w_{i,Y}\boldsymbol{\alpha}_{Y,k}^\top\boldsymbol{\Sigma}_{Y,k}^{-1} \boldsymbol{\alpha}_{Y,k}\\
&+\frac{1}{2}\sum_{i=1}^n\sum_{k=1}^K z_{ik}\biggl(\mathbfit{c}_{Y,i}^\top\boldsymbol{\Sigma}_{Y,k}^{-1}\boldsymbol{\alpha}_{Y,k}-\left({\mathbfit{c}_{X,i}^{*}}\right)^\top\left(\boldsymbol{\Gamma}^k_{*}\right)^\top\boldsymbol{\Sigma}_{Y,k}^{-1}\boldsymbol{\alpha}_{Y,k}+\boldsymbol{\alpha}_{Y,k}^\top\boldsymbol{\Sigma}_{Y,k}^{-1}\mathbfit{c}_{Y,i}-\boldsymbol{\alpha}_{Y,k}^\top\boldsymbol{\Sigma}_{Y,k}^{-1}\boldsymbol{\Gamma}^k_{*}{\mathbfit{c}_{X,i}^{*}}\biggl)
\end{align*}
\end{proof}
\section{Proof of Proposition \ref{propEM}}
\label{secA2}
\begin{proof}
Similarly with \eqref{delty} we obtain
\begin{align}
&(\mathbfit{c}_{X,i}-\boldsymbol{\mu}_{X,k})^\top\boldsymbol{\Sigma}_{X,k}^{-1}(\mathbfit{c}_{X,i}-\boldsymbol{\mu}_{X,k})=\biggl(\sum_{l=1}^{d_k}\frac{\mathbfit{q}_{kl}^\top \mathbfit{W}_X^{1/2}(\mathbfit{c}_{X,i}-\boldsymbol{\mu}_{X,k})(\mathbfit{c}_{X,i}-\boldsymbol{\mu}_{X,k})^\top \mathbfit{W}_X^{1/2}\mathbfit{q}_{kl}}{a_{kl}}\nonumber\\
&+\sum_{l=d_k+1}^{R}\frac{\mathbfit{q}_{kl}^\top \mathbfit{W}_X^{1/2}(\mathbfit{c}_{X,i}-\boldsymbol{\mu}_{X,k})(\mathbfit{c}_{X,i}-\boldsymbol{\mu}_{X,k})^\top\mathbfit{W}_X^{1/2}\mathbfit{q}_{kl}}{b_{k}}\biggl)=\delta_{X,k}(\mathbfit{c}_{X,i})\nonumber,\\
&\boldsymbol{\alpha}_{X,k}^\top\boldsymbol{\Sigma}_{X,k}^{-1}\boldsymbol{\alpha}_{X,k}=\biggl(\sum_{l=1}^{d_k}\frac{\mathbfit{q}_{kl}^\top \mathbfit{W}_X^{1/2}\boldsymbol{\alpha}_{X,k}\boldsymbol{\alpha}_{X,k}^\top \mathbfit{W}_X^{1/2}\mathbfit{q}_{kl}}{a_{kl}}+\sum_{l=d_k+1}^{R}\frac{\mathbfit{q}_{kl}^\top \mathbfit{W}_X^{1/2}\boldsymbol{\alpha}_{X,k}\boldsymbol{\alpha}_{X,k}^\top\mathbfit{W}_X^{1/2}\mathbfit{q}_{kl}}{b_{k}}\biggl)=\rho_{X,k}\nonumber.\\
&(\mathbfit{c}_{X,i}-\boldsymbol{\mu}_{X,k})^\top \boldsymbol{\Sigma}_{X,k}^{-1}\boldsymbol{\alpha}_{X,k}=\biggl(\sum_{l=1}^{d_k}\frac{\mathbfit{q}_{kl}^\top \mathbfit{W}_X^{1/2}\boldsymbol{\alpha}_{X,k}(\mathbfit{c}_{X,i}-\boldsymbol{\mu}_{X,k})^\top \mathbfit{W}_X^{1/2}\mathbfit{q}_{kl}}{a_{kl}}\nonumber\\
&+\sum_{l=d_k+1}^{R}\frac{\mathbfit{q}_{kl}^\top \mathbfit{W}_X^{1/2}\boldsymbol{\alpha}_{X,k}(\mathbfit{c}_{X,i}-\boldsymbol{\mu}_{X,k})^\top\mathbfit{W}_X^{1/2}\mathbfit{q}_{kl}}{b_{k}}\biggl)\nonumber=\delta_{\alpha,k}(\mathbfit{c}_{X,i}).\\
\end{align}
Replacing in \eqref{dens} and using also \eqref{pdfUNIF} and \eqref{dety} we obtain
\begin{align*}
&p_k(\mathbfit{c}_{Y,i},\mathbfit{c}_{X,i} \mid \boldsymbol{\theta}_k)=f_{SKW}(\mathbfit{c}_{X,i};\boldsymbol{\mu}_{X,k},\boldsymbol{\alpha}_{X,k},\boldsymbol{\Sigma}_{X,k},p_{X,1,k},p_{X,2,k},p_{X,3,k},p_{X,4,k})\\
&f_{SKW}(\mathbfit{c}_{Y,i};\boldsymbol{\mu}_{Y,k},\boldsymbol{\alpha}_{Y,k},\boldsymbol{\Sigma}_{Y,k},p_{Y,1,k},p_{Y,2,k},p_{Y,3,k},p_{Y,4,k})\\
&=(2\pi)^{-(R_X+R_Y)/2}\mid \boldsymbol{\Sigma}_{X,k}\mid ^{-1/2}\mid\boldsymbol{\Sigma}_{Y,k}\mid ^{-1/2}\exp\biggl(\delta_{\alpha,k}(\mathbfit{c}_{X,i})+p_{X,3,k}\log\left(\delta_{X,k}(\mathbfit{c}_{X,i})+p_{X,1,k}\right)\\
&-p_{X,3,k}\log\left(\rho_{X,k}+p_{X,2,k}\right)+p_{X,4,k}+\log\left( K_{2p_{X,3,k}}\left(\sqrt{\left(\rho_{X,k}+p_{X,2,k}\right)\left(\delta_{X,k}(\mathbfit{c}_{X,i})+p_{X,1,k}\right)}\right)\right)\\
&+\left(\mathbfit{c}_{Y,i}-\boldsymbol{\mu}_{Y,k}\right)^\top\boldsymbol{\Sigma}_{Y,k}^{-1}\boldsymbol{\alpha}_{Y,k}+p_{Y,3,k}\log\left(\delta(\mathbfit{c}_{Y,i};\boldsymbol{\mu}_{Y,k}, \boldsymbol{\Sigma}_{Y,k})+p_{Y,1,k}\right)\notag\\
&-p_{Y,3,k}\log\left(\rho(\boldsymbol{\alpha}_{Y,k},\boldsymbol{\Sigma}_{Y,k})+p_{Y,2,k}\right)+p_{Y,4,k}\\&+\log\left( K_{2p_{Y,3,k}}\left(\sqrt{\left(\rho(\boldsymbol{\alpha}_{Y,k},\boldsymbol{\Sigma}_{Y,k})+p_{Y,2,k}\right)\left(\delta(\mathbfit{c}_{Y,i};\boldsymbol{\mu}_{Y,k}, \boldsymbol{\Sigma}_{Y,k})+p_{Y,1,k}\right)}\right)\right)\biggl)\\
&=(2\pi)^{-(R_X+R_Y)/2}\left(\prod_{j=1}^{d_k} a_{kj}\prod_{j=d_k+1}^{R_X}b_k\right)^{-1/2}\mid \mathbfit{W}_X\mid ^{1/2}\mid\boldsymbol{\Sigma}_{Y,k}\mid ^{-1/2}\\
&\exp\biggl(\delta_{\alpha,k}(\mathbfit{c}_{X,i})+p_{X,3,k}\log\left(\delta_{X,k}(\mathbfit{c}_{X,i})+p_{X,1,k}\right)-p_{X,3,k}\log\left(\rho_{X,k}+p_{X,2,k}\right)+p_{X,4,k}\\
&+\log\left( K_{2p_{X,3,k}}\left(\sqrt{\left(\rho_{X,k}+p_{X,2,k}\right)\left(\delta_{X,k}(\mathbfit{c}_{X,i})+p_{X,1,k}\right)}\right)\right)\\
&+\left(\mathbfit{c}_{Y,i}-\boldsymbol{\Gamma}^k_{*}{\mathbfit{c}_{X,i}^{*}}\right)^\top\boldsymbol{\Sigma}_{Y,k}^{-1}\boldsymbol{\alpha}_{Y,k}+p_{Y,3,k}\log\left(\delta(\mathbfit{c}_{Y,i};\boldsymbol{\Gamma}^k_{*}{\mathbfit{c}_{X,i}^{*}}, \boldsymbol{\Sigma}_{Y,k})+p_{Y,1,k}\right)\notag\\
&-p_{Y,3,k}\log\left(\rho(\boldsymbol{\alpha}_{Y,k},\boldsymbol{\Sigma}_{Y,k})+p_{Y,2,k}\right)+p_{Y,4,k}\\
&+\log\left( K_{2p_{Y,3,k}}\left(\sqrt{\left(\rho(\boldsymbol{\alpha}_{Y,k},\boldsymbol{\Sigma}_{Y,k})+p_{Y,2,k}\right)\left(\delta(\mathbfit{c}_{Y,i};\boldsymbol{\Gamma}^k_{*}{\mathbfit{c}_{X,i}^{*}}, \boldsymbol{\Sigma}_{Y,k})+p_{Y,1,k}\right)}\right)\right)\biggl)\\
&=(2\pi)^{-(R_X+R_Y)/2}\mid \mathbfit{W}_X\mid ^{1/2}\exp\biggl(-\frac{1}{2}\biggl(\sum_{j=1}^{d_k} \log (a_{kj})\\
&+(R_X-d_k)\log (b_k)+\log(\mid\boldsymbol{\Sigma}_{Y,k}\mid)\biggl)+\delta_{\alpha,k}(\mathbfit{c}_{X,i})\\
&+p_{X,3,k}\log\left(\delta_{X,k}(\mathbfit{c}_{X,i})+p_{X,1,k}\right)-p_{X,3,k}\log\left(\rho_{X,k}+p_{X,2,k}\right)+p_{X,4,k}\\
&+\log\left( K_{2p_{X,3,k}}\left(\sqrt{\left(\rho_{X,k}+p_{X,2,k}\right)\left(\delta_{X,k}(\mathbfit{c}_{X,i})+p_{X,1,k}\right)}\right)\right)\\
&+\left(\mathbfit{c}_{Y,i}-\boldsymbol{\Gamma}^k_{*}{\mathbfit{c}_{X,i}^{*}}\right)^\top\boldsymbol{\Sigma}_{Y,k}^{-1}\boldsymbol{\alpha}_{Y,k}+p_{Y,3,k}\log\left(\delta(\mathbfit{c}_{Y,i};\boldsymbol{\Gamma}^k_{*}{\mathbfit{c}_{X,i}^{*}}, \boldsymbol{\Sigma}_{Y,k})+p_{Y,1,k}\right)\notag\\
&-p_{Y,3,k}\log\left(\rho(\boldsymbol{\alpha}_{Y,k},\boldsymbol{\Sigma}_{Y,k})+p_{Y,2,k}\right)+p_{Y,4,k}\\
&+\log\left( K_{2p_{Y,3,k}}\left(\sqrt{\left(\rho(\boldsymbol{\alpha}_{Y,k},\boldsymbol{\Sigma}_{Y,k})+p_{Y,2,k}\right)\left(\delta(\mathbfit{c}_{Y,i};\boldsymbol{\Gamma}^k_{*}{\mathbfit{c}_{X,i}^{*}}, \boldsymbol{\Sigma}_{Y,k})+p_{Y,1,k}\right)}\right)\right)\biggl)\\
&=(2\pi)^{-(R_X+R_Y)/2}\mid \mathbfit{W}_X\mid ^{1/2}\pi_k^{-1}\exp\left(H_k(\mathbfit{c}_{Y,i},\mathbfit{c}_{X,i}\mid \boldsymbol{\theta}_k)\right),
\end{align*}
where $H_k(\mathbfit{c}_{Y,i},\mathbfit{c}_{X,i}\mid \boldsymbol{\theta}_k)$ is defined in \eqref{Hk}.
\end{proof}
\section{Proof of Proposition \ref{propM}}
\label{secA3}
\begin{proof}
Using \eqref{lik2}-\eqref{l4c2} we have that $Q(\boldsymbol{\theta}\mid \boldsymbol{\theta}^{(m-1)})$ is given by
\begin{equation*}
Q(\boldsymbol{\theta}\mid \boldsymbol{\theta}^{(m-1)})=Q_1(\pi\mid \boldsymbol{\theta}^{(m-1)})+Q_2(\boldsymbol{\vartheta}_X\mid \boldsymbol{\theta}^{(m-1)})+Q_3(\boldsymbol{\vartheta}_Y\mid \boldsymbol{\theta}^{(m-1)})+Q_4(\theta_{W,X}\cup \theta_{W,Y}\mid \boldsymbol{\theta}^{(m-1)}),
\end{equation*}
were 
\begin{align*}
&Q_1(\pi\mid \boldsymbol{\theta}^{(m-1)})=\sum_{i=1}^n\sum_{k=1}^K t_{ik}^{(m)}\log(\pi_k)\\
&Q_2(\boldsymbol{\vartheta}_X\mid \boldsymbol{\theta}^{(m-1)})=-\frac{1}{2}\sum_{i=1}^n\sum_{k=1}^K t_{ik}^{(m)}lw_{ik,X}^{(m)}-\frac{nR_X\log(2\pi)}{2}+\frac{n}{2}\log(\mid \mathbfit{W}_X\mid)\\
&-\frac{1}{2}\sum_{k=1}^K n_k^{(m)}\sum_{l=1}^{d_k}\log(a_{kl})-\frac{1}{2}\sum_{k=1}^K n_k^{(m)}\sum_{l=d_k+1}^{R_X}\log(b_{k})\\
&-\frac{1}{2}\sum_{k=1}^K n_k^{(m)}  \biggl(\sum_{l=1}^{d_k}\frac{\mathbfit{q}_{kl}^\top \mathbfit{W}_X^{1/2}\mathbfit{S}_{X,k}^{(m)}\mathbfit{W}_X^{1/2}\mathbfit{q}_{kl}}{a_{kl}}+\sum_{l=d_k+1}^{R_X}\frac{\mathbfit{q}_{kl}^\top \mathbfit{W}_X^{1/2}\mathbfit{S}_{X,k}^{(m)}\mathbfit{W}_X^{1/2}\mathbfit{q}_{kl}}{b_{k}}\biggl),\\
&Q_3(\boldsymbol{\vartheta}_Y\mid \boldsymbol{\theta}^{(m-1)})=-\frac{1}{2}\sum_{i=1}^n\sum_{k=1}^K t_{ik}^{(m)}lw_{ik,Y}^{(m)}-\frac{nR_Y\log(2\pi)}{2}-\frac{1}{2}\sum_{k=1}^K n_{k}^{(m)}\log(\mid \boldsymbol{\Sigma}_{Y,k}\mid) \\
&-\frac{1}{2}\sum_{i=1}^n\sum_{k=1}^K t_{ik}^{(m)}wi_{ik,Y}^{(m)}(\mathbfit{c}_{Y,i}-\boldsymbol{\Gamma}^k_{*}{\mathbfit{c}_{X,i}^{*}})^\top\boldsymbol{\Sigma}_{Y,k}^{-1}
(\mathbfit{c}_{Y,i}-\boldsymbol{\Gamma}^k_{*}{\mathbfit{c}_{X,i}^{*}})\\
&-\frac{1}{2}\sum_{i=1}^n\sum_{k=1}^K t_{ik}^{(m)}w_{ik,Y}^{(m)}\boldsymbol{\alpha}_{Y,k}^\top\boldsymbol{\Sigma}_{Y,k}^{-1}\boldsymbol{\alpha}_{Y,k}\\
&+\frac{1}{2}\sum_{i=1}^n\sum_{k=1}^K t_{ik}^{(m)}\left(\left(\mathbfit{c}_{Y,i}
-\boldsymbol{\Gamma}^k_{*}{\mathbfit{c}_{X,i}^{*}}\right)^\top\boldsymbol{\Sigma}_{Y,k}^{-1}\boldsymbol{\alpha}_{Y,k}
+\boldsymbol{\alpha}_{Y,k}^\top\boldsymbol{\Sigma}_{Y,k}^{-1}\left(\mathbfit{c}_{Y,i}-\boldsymbol{\Gamma}^k_{*}{\mathbfit{c}_{X,i}^{*}}\right)\right)\\
&Q_4(\theta_{W,X}\cup \theta_{W,Y}\mid \boldsymbol{\theta}^{(m-1)})=\sum_{i=1}^n\sum_{k=1}^KE[ z_{ik}\log (h(w_{i,X};\theta_{W,X}))\mid \mathbfit{c}_{X,1},\ldots,\mathbfit{c}_{X,n},\boldsymbol{\theta}^{(m-1)} ]\\
&\sum_{i=1}^n\sum_{k=1}^KE[ z_{ik}\log (h(w_{i,Y};\theta_{W,Y}))\mid \mathbfit{c}_{Y,1},\mathbfit{c}_{X,1},\ldots,\mathbfit{c}_{Y,n},\mathbfit{c}_{X,n},\boldsymbol{\theta}^{(m-1)}]
\end{align*}
where $\mathbfit{S}_{X,k}^{(m)}$ is defined in \eqref{sig}. The formulas for $Q_4(\theta_{W,X}\cup \theta_{W,Y}\mid \boldsymbol{\theta}^{(m-1)})$ depend on the specific pair of distributions.

For the estimation of $\pi_k$, $k=1,\ldots, K$ we introduce the Lagrange multiplier $\lambda$ and we maximize 
$Q_1=Q_1(\pi\mid \boldsymbol{\theta}^{(m-1)})-\lambda(\sum_{k=1}^K \pi_k-1)$. We get \eqref{tpi} solving the system
\begin{equation*}
\frac{\partial Q_1}{\partial \pi_k}=\sum_{i=1}^n\frac{ t_{ik}^{(m)}}{\pi_k}-\lambda=0, k=1,\ldots, K\quad \frac{\partial Q_1}{\partial \lambda}=\sum_{k=1}^K \pi_k-1=0.
\end{equation*}

To get an update for $\boldsymbol{\boldsymbol{\mu}}_{X,k}^{(m)}$ and $\boldsymbol{\boldsymbol{\alpha}}_{X,k}^{(m)}$ we start start from the formula \eqref{l3use}:
\begin{align*}
&Q_2(\boldsymbol{\vartheta}_X\mid \boldsymbol{\theta}^{(m-1)})=-\frac{n}{2}R_X\log(2\pi)-\frac{1}{2}\sum_{k=1}^K n_{k}^{(m)}\log\mid \boldsymbol{\Sigma}_{X,k}\mid-\frac{1}{2}\sum_{i=1}^n\sum_{k=1}^K t_{ik}^{(m)}lw_{ik,X}^{(m)} \nonumber\\
&-\frac{1}{2}\sum_{i=1}^n\sum_{k=1}^K t_{ik}^{(m)}wi_{ik,X}^{(m)}(\mathbfit{c}_{X,i}-\boldsymbol{\boldsymbol{\mu}}_{X,k})^\top\boldsymbol{\Sigma}_{X,k}^{-1}(\mathbfit{c}_{X,i}-\boldsymbol{\boldsymbol{\mu}}_{X,k})-\frac{1}{2}\sum_{i=1}^n\sum_{k=1}^K t_{ik}^{(m)}w_{ik,X}^{(m)}\boldsymbol{\alpha}_{X,k}^\top\boldsymbol{\Sigma}_{X,k}^{-1} \boldsymbol{\alpha}_{X,k}\\
&+\frac{1}{2}\sum_{i=1}^n\sum_{k=1}^K t_{ik}^{(m)}\left((\mathbfit{c}_{X,i}-\boldsymbol{\mu}_{X,k})^\top\boldsymbol{\Sigma}_{X,k}^{-1}\boldsymbol{\alpha}_{X,k}+\boldsymbol{\alpha}_{X,k}^\top\boldsymbol{\Sigma}_{X,k}^{-1}(\mathbfit{c}_{X,i}-\boldsymbol{\mu}_{X,k})\right).
\end{align*}
The gradient of $Q_2$ with respect to  ${\boldsymbol{\mu}}_{X,k}$ is
\begin{align*}
&\nabla_{{\boldsymbol{\mu}}_{X,k}} Q_2(\boldsymbol{\vartheta}_X\mid \boldsymbol{\theta}^{(m-1)})=\sum_{i=1}^n t_{ik}^{(m)}wi_{ik,X}^{(m)}\boldsymbol{\Sigma}_{X,k}^{-1}(\mathbfit{c}_{X,i}-\boldsymbol{\mu}_{X,k})-\sum_{i=1}^n t_{ik}^{(m)}\boldsymbol{\Sigma}_{X,k}^{-1}\boldsymbol{\alpha}_{X,k}\nonumber\\
&=\boldsymbol{\Sigma}_{X,k}^{-1}\left(\sum_{i=1}^n t_{ik}^{(m)}wi_{ik,X}^{(m)}\mathbfit{c}_{X,i}- n_{k}^{(m)}\boldsymbol{\alpha}_{X,k}- \sum_{i=1}^n t_{ik}^{(m)}wi_{ik,X}^{(m)}{\boldsymbol{\mu}}_{X,k}\right).
\end{align*}
The gradient of $Q_2$ with respect to  ${\boldsymbol{\alpha}}_{X,k}$ is
\begin{align*}
&\nabla_{{\boldsymbol{\alpha}}_{X,k}} Q_2(\boldsymbol{\vartheta}_X\mid \boldsymbol{\theta}^{(m-1)})=-\sum_{i=1}^n t_{ik}^{(m)}w_{ik,X}^{(m)}\boldsymbol{\Sigma}_{X,k}^{-1}\boldsymbol{\alpha}_{X,k}+\sum_{i=1}^n t_{ik}^{(m)}\boldsymbol{\Sigma}_{X,k}^{-1}(\mathbfit{c}_{X,i}-\boldsymbol{\mu}_{X,k})\nonumber\\
&=\boldsymbol{\Sigma}_{X,k}^{-1}\left(\sum_{i=1}^n t_{ik}^{(m)}\mathbfit{c}_{X,i}- n_{k}^{(m)}{\boldsymbol{\mu}}_{X,k}-\sum_{i=1}^n t_{ik}^{(m)}w_{ik,X}^{(m)}\boldsymbol{\alpha}_{X,k}\right).
\end{align*}

Solving $\nabla_{{\boldsymbol{\mu}}_{X,k}} Q_2(\boldsymbol{\vartheta}_X\mid \boldsymbol{\theta}^{(m-1)})={\bf 0}$, $\nabla_{{\boldsymbol{\alpha}}_{X,k}} Q_2(\boldsymbol{\vartheta}_X\mid \boldsymbol{\theta}^{(m-1)})={\bf 0}$ we get formulas \eqref{tmu} and \eqref{formalpha}.

To estimate $\mathbfit{Q}_k$ we have to maximize $Q_2(\boldsymbol{\vartheta}_X\mid \boldsymbol{\theta}^{(m-1)})$ with respect to $\mathbfit{q}_{kl}$ under the constraint   $\mathbfit{q}_{kl}^\top\mathbfit{q}_{kl}=1$. This is equivalent with minimizing $-2Q_2(\boldsymbol{\vartheta}_X\mid \boldsymbol{\theta}^{(m-1)})$ with respect to $\mathbfit{q}_{kl}$ under this constraint, so  we consider the  function $Q_{2c}=-2Q_2(\boldsymbol{\vartheta}_X\mid \boldsymbol{\theta}^{(m-1)})-\sum_{l=1}^{R_X} \omega_{kl}(\mathbfit{q}_{kl}^\top\mathbfit{q}_{kl}-1)$, where $\omega_{kl}$ are Lagrange multipliers.
The gradient of $Q_{2c}$ with respect to $\mathbfit{q}_{kl}$ is
\begin{align*}
\nabla_{\mathbfit{q}_{kl}}Q_{2c}&= 2n_k^{(m)}\frac{ \mathbfit{W}_X^{1/2}\mathbfit{S}_{X,k}^{(m)}\mathbfit{W}_X^{1/2}\mathbfit{q}_{kl}}{\Sigma_{kl}}-2\omega_{kl}\mathbfit{q}_{kl},\\
&\Sigma_{kl}=\begin{cases}&a_{kl}\text{ if } l=1,\ldots, d_k\\
&b_k\text{ if } l=d_k+1,\ldots, R_X.
\end{cases}
\end{align*}
From $\nabla_{\mathbfit{q}_{kl}}Q_{2c}=0$ we get $ \mathbfit{W}_X^{1/2}\mathbfit{S}_{X,k}^{(m)}\mathbfit{W}_X^{1/2}\mathbfit{q}_{kl}=\frac{\omega_{kl}\Sigma_{kl}}{n_k^{(m)}}\mathbfit{q}_{kl}$, so $\mathbfit{q}_{kl}$ is an eigenfunction of $\mathbfit{W}_X^{1/2}\mathbfit{S}_{X,k}^{(m)}\mathbfit{W}_X^{1/2}$ and the associated eigenvalue is $\lambda_{kl}^{(m)}=\frac{\omega_{kl}\Sigma_{kl}}{n_k^{(m)}}$. Notice that we also have $\mathbfit{q}_{kl}^\top\mathbfit{q}_{kj}=0$ if $l\ne j$, and  $\lambda_{kl}^{(m)}=\mathbfit{q}_{kl}^\top\mathbfit{W}_X^{1/2}\mathbfit{S}_{X,k}^{(m)}\mathbfit{W}_X^{1/2}\mathbfit{q}_{kl}$ so we can write
\begin{align*}
&-2Q_2(\boldsymbol{\vartheta}_X\mid \boldsymbol{\theta}^{(m-1)})=nR_X\log(2\pi)-n\log(\mid \mathbfit{W}_X\mid)+\sum_{k=1}^K n_k^{(m)}\biggl(\sum_{l=1}^{d_k}\log(a_{kl})\\
&+\sum_{l=d_k+1}^{R_X}\log(b_{k})\biggl)+\sum_{k=1}^K n_k^{(m)}  \biggl(\sum_{l=1}^{d_k}\frac{\lambda_{kl}^{(m)}}{a_{kl}}+\sum_{l=d_k+1}^{R_X}\frac{\lambda_{kl}^{(m)}}{b_{k}}\biggl)+\sum_{i=1}^n\sum_{k=1}^K t_{ik}^{(m)}lw_{ik,X}^{(m)} \\
&=nR_X\log(2\pi)-n\log(\mid \mathbfit{W}_X\mid)+\sum_{k=1}^K n_k^{(m)}\biggl(\sum_{l=1}^{d_k}\log(a_{kl})+\sum_{l=d_k+1}^{R_X}\log(b_{k})\biggl)\\
&+\sum_{k=1}^K n_k^{(m)}  \biggl(\sum_{l=1}^{d_k}\lambda_{kl}^{(m)}\left(\frac{1}{a_{kl}}-\frac{1}{b_k}\right)+\frac{1}{b_k} \text{trace}(\mathbfit{W}_X^{1/2}\mathbfit{S}_{X,k}^{(m)}\mathbfit{W}_X^{1/2})\biggl)+\sum_{i=1}^n\sum_{k=1}^K t_{ik}^{(m)}lw_{ik,X}^{(m)} .
\end{align*}
Here we have also used 
\begin{equation}
\text{trace}(\mathbfit{W}_X^{1/2}\mathbfit{S}_{X,k}^{(m)}\mathbfit{W}_X^{1/2})=\sum_{l=1}^{R_X} \lambda_{kl}^{(m)}=\sum_{l=1}^{d_k} \lambda_{kl}^{(m)}+\sum_{l=d_k+1}^{R_X} \lambda_{kl}^{(m)}.\label{trac}
\end{equation}
Since for any $l=1,\ldots, d_k$ we have $a_{kl}\ge b_k$, we get $\frac{1}{a_{kl}}-\frac{1}{b_k}\le 0$, so $\sum_{l=1}^{d_k}\lambda_{kl}^{(m)}\left(\frac{1}{a_{kl}}-\frac{1}{b_k}\right)$ is a decreasing function of $\lambda_{kl}$. Thus,  we estimate $\mathbfit{q}_{kl}$ by the eigenfunction associated with the $l$th highest eigenvalue of $\mathbfit{W}_X^{1/2}\mathbfit{S}_{X,k}^{(m)}\mathbfit{W}_X^{1/2}$.

To update $a_{kl}$ we solve
\begin{align*}
\frac{\partial Q_2(\boldsymbol{\vartheta}_X\mid \boldsymbol{\theta}^{(m-1)})}{\partial a_{kl}}=-\frac{n_k^{(m)}}{2a_{kl}}+\frac{n_k^{(m)}\mathbfit{q}_{kl}^\top \mathbfit{W}^{1/2}\mathbfit{S}_{X,k}^{(m)}\mathbfit{W}^{1/2}\mathbfit{q}_{kl}}{2a_{kl}^2}=0,
\end{align*}
and we get $a_{kl}^{(m)}=\mathbfit{q}_{kl}^\top \mathbfit{W}_X^{1/2}\mathbfit{S}_{X,k}^{(m)}\mathbfit{W}_X^{1/2}\mathbfit{q}_{kl}=\lambda_{kl}^{(m)}$, the $l$th highest eigenvalue of $\mathbfit{W}_X^{1/2}\mathbfit{S}_{X,k}^{(m)}\mathbfit{W}_X^{1/2}$.

From 
\begin{align*}
\frac{\partial Q_2(\boldsymbol{\vartheta}_X\mid \boldsymbol{\theta}^{(m-1)})}{\partial b_{k}}=-\frac{n_k^{(m)}}{2}\sum_{l=d_k+1}^{R_X}\frac{1}{b_{k}}+\frac{n_k^{(m)}}{2}\sum_{l=d_k+1}^{R_X}\frac{\mathbfit{q}_{kl}^\top \mathbfit{W}_X^{1/2}\mathbfit{S}_{X,k}^{(m)}\mathbfit{W}_X^{1/2}\mathbfit{q}_{kl}}{b_{k}^2}=0,
\end{align*}
we obtain
$$
b_{k}^{(m)}=\frac{1}{R_X-d_k}\sum_{l=d_k+1}^{R_X}\mathbfit{q}_{kl}^\top \mathbfit{W}_X^{1/2}\mathbfit{S}_{X,k}^{(m)}\mathbfit{W}_X^{1/2}\mathbfit{q}_{kl}=\frac{1}{R_X-d_k}\sum_{l=d_k+1}^{R_X}\lambda_{kl}^{(m)}
$$
Thus,  using \eqref{trac} we get
$$
b_{k}^{(m)}=\frac{1}{R_X-d_k}\left(\text{trace}(\mathbfit{W}_X^{1/2}\mathbfit{S}_{X,k}^{(m)}\mathbfit{W}_X^{1/2})-
\sum_{l=1}^{d_k} \lambda_{kl}^{(m)}\right).
$$ 
To estimate the regression coefficient we use the properties of trace and transpose and $\boldsymbol{\Sigma}_{Y,k}^{\top}=\boldsymbol{\Sigma}_{Y,k}$  and we get
\begin{align*}
&Q_3(\boldsymbol{\vartheta}_Y\mid \boldsymbol{\theta}^{(m-1)})=-\frac{1}{2}\sum_{i=1}^n\sum_{k=1}^K t_{ik}^{(m)}lw_{ik,Y}^{(m)}-\frac{nR_Y\log(2\pi)}{2}-\frac{1}{2}\sum_{k=1}^K n_{k}^{(m)}\log(\mid \boldsymbol{\Sigma}_{Y,k}\mid) \\
&-\frac{1}{2}\sum_{i=1}^n\sum_{k=1}^K t_{ik}^{(m)}wi_{ik,Y}^{(m)}\biggl(\mathbfit{c}_{Y,i}^\top\boldsymbol{\Sigma}_{Y,k}^{-1}\mathbfit{c}_{Y,i}-\text{trace}\left(\mathbfit{c}_{Y,i}^\top\boldsymbol{\Sigma}_{Y,k}^{-1}\boldsymbol{\Gamma}^k_{*}{\mathbfit{c}_{X,i}^{*}}\right)\\
&-\text{trace}\left(({\mathbfit{c}_{X,i}^{*}})^\top\left(\boldsymbol{\Gamma}^k_{*}\right)^\top \boldsymbol{\Sigma}_{Y,k}^{-1}\mathbfit{c}_{Y,i}\right)+\text{trace}\left(({\mathbfit{c}_{X,i}^{*}})^\top\left(\boldsymbol{\Gamma}^k_{*}\right)^\top\boldsymbol{\Sigma}_{Y,k}^{-1}\boldsymbol{\Gamma}^k_{*}{\mathbfit{c}_{X,i}^{*}}\right)\biggl)\\
&-\frac{1}{2}\sum_{i=1}^n\sum_{k=1}^K t_{ik}^{(m)}w_{ik,Y}^{(m)}\text{trace}\left(\boldsymbol{\alpha}_{Y,k}^\top\boldsymbol{\Sigma}_{Y,k}^{-1}\boldsymbol{\alpha}_{Y,k}\right)\\
&+\frac{1}{2}\sum_{i=1}^n\sum_{k=1}^K t_{ik}\biggl(\text{trace}\left(\mathbfit{c}_{Y,i}^\top\boldsymbol{\Sigma}_{Y,k}^{-1}\boldsymbol{\alpha}_{Y,k}\right)-\text{trace}\left(\left({\mathbfit{c}_{X,i}^{*}}\right)^\top\left(\boldsymbol{\Gamma}^k_{*}\right)^\top\boldsymbol{\Sigma}_{Y,k}^{-1}\boldsymbol{\alpha}_{Y,k}\right)\\
&+\text{trace}\left(\boldsymbol{\alpha}_{Y,k}^\top\boldsymbol{\Sigma}_{Y,k}^{-1}\mathbfit{c}_{Y,i}\right)-\text{trace}\left(\boldsymbol{\alpha}_{Y,k}^\top\boldsymbol{\Sigma}_{Y,k}^{-1}\boldsymbol{\Gamma}^k_{*}{\mathbfit{c}_{X,i}^{*}}\right)\biggl)\\
&=-\frac{1}{2}\sum_{i=1}^n\sum_{k=1}^K t_{ik}^{(m)}lw_{ik,Y}^{(m)}-\frac{nR_Y\log(2\pi)}{2}-\frac{1}{2}\sum_{k=1}^K n_{k}^{(m)}\log(\mid \boldsymbol{\Sigma}_{Y,k}\mid)\\
&-\frac{1}{2}\sum_{i=1}^n\sum_{k=1}^K t_{ik}^{(m)}wi_{ik,Y}^{(m)}\biggl(\mathbfit{c}_{Y,i}^\top\boldsymbol{\Sigma}_{Y,k}^{-1}\mathbfit{c}_{Y,i}-\text{trace}\left(\boldsymbol{\Gamma}^k_{*}{\mathbfit{c}_{X,i}^{*}}\mathbfit{c}_{Y,i}^\top\boldsymbol{\Sigma}_{Y,k}^{-1}\right)\\
&-\text{trace}\left(\boldsymbol{\Sigma}_{Y,k}^{-1}\mathbfit{c}_{Y,i}({\mathbfit{c}_{X,i}^{*}})^\top\left(\boldsymbol{\Gamma}^k_{*}\right)^\top \right)+\text{trace}\left(\boldsymbol{\Gamma}^k_{*}{\mathbfit{c}_{X,i}^{*}}({\mathbfit{c}_{X,i}^{*}})^\top\left(\boldsymbol{\Gamma}^k_{*}\right)^\top\boldsymbol{\Sigma}_{Y,k}^{-1}\right)\biggl)\\
&-\frac{1}{2}\sum_{i=1}^n\sum_{k=1}^K t_{ik}^{(m)}w_{ik,Y}^{(m)}\text{trace}\left(\boldsymbol{\alpha}_{Y,k}\boldsymbol{\alpha}_{Y,k}^\top\boldsymbol{\Sigma}_{Y,k}^{-1}\right)\\
&+\frac{1}{2}\sum_{i=1}^n\sum_{k=1}^K t_{ik}\biggl(\text{trace}\left(\boldsymbol{\alpha}_{Y,k}\mathbfit{c}_{Y,i}^\top\boldsymbol{\Sigma}_{Y,k}^{-1}\right)-\text{trace}\left(\boldsymbol{\alpha}_{Y,k}\left({\mathbfit{c}_{X,i}^{*}}\right)^\top\left(\boldsymbol{\Gamma}^k_{*}\right)^\top\boldsymbol{\Sigma}_{Y,k}^{-1}\right)\\
&+\text{trace}\left(\boldsymbol{\Sigma}_{Y,k}^{-1}\mathbfit{c}_{Y,i}\boldsymbol{\alpha}_{Y,k}^\top\right)-\text{trace}\left(\boldsymbol{\Sigma}_{Y,k}^{-1}\boldsymbol{\Gamma}^k_{*}{\mathbfit{c}_{X,i}^{*}}\boldsymbol{\alpha}_{Y,k}^\top\right)\biggl)\\
&=-\frac{1}{2}\sum_{i=1}^n\sum_{k=1}^K t_{ik}^{(m)}lw_{ik,Y}^{(m)}-\frac{nR_Y\log(2\pi)}{2}-\frac{1}{2}\sum_{k=1}^K n_{k}^{(m)}\log(\mid \boldsymbol{\Sigma}_{Y,k}\mid)\\
&-\frac{1}{2}\sum_{i=1}^n\sum_{k=1}^K t_{ik}^{(m)}wi_{ik,Y}^{(m)}\biggl(\mathbfit{c}_{Y,i}^\top\boldsymbol{\Sigma}_{Y,k}^{-1}\mathbfit{c}_{Y,i}-2\text{trace}\left(\boldsymbol{\Gamma}^k_{*}{\mathbfit{c}_{X,i}^{*}}\mathbfit{c}_{Y,i}^\top\boldsymbol{\Sigma}_{Y,k}^{-1}\right)\\
&+\text{trace}\left(\boldsymbol{\Gamma}^k_{*}{\mathbfit{c}_{X,i}^{*}}({\mathbfit{c}_{X,i}^{*}})^\top\left(\boldsymbol{\Gamma}^k_{*}\right)^\top\boldsymbol{\Sigma}_{Y,k}^{-1}\right)\biggl)\\
&-\frac{1}{2}\sum_{i=1}^n\sum_{k=1}^K t_{ik}^{(m)}w_{ik,Y}^{(m)}\text{trace}\left(\boldsymbol{\alpha}_{Y,k}\boldsymbol{\alpha}_{Y,k}^\top\boldsymbol{\Sigma}_{Y,k}^{-1}\right)\\
&+\sum_{i=1}^n\sum_{k=1}^K t_{ik}\biggl(\text{trace}\left(\boldsymbol{\alpha}_{Y,k}\mathbfit{c}_{Y,i}^\top\boldsymbol{\Sigma}_{Y,k}^{-1}\right)-\text{trace}\left(\boldsymbol{\alpha}_{Y,k}\left({\mathbfit{c}_{X,i}^{*}}\right)^\top\left(\boldsymbol{\Gamma}^k_{*}\right)^\top\boldsymbol{\Sigma}_{Y,k}^{-1}\right)
\end{align*}
To update $\boldsymbol{\Gamma}^k_{*}$ and $\boldsymbol{\alpha}_{Y,k}$  and get formulas \eqref{GammY} and \eqref{alphY} we solve
\begin{align*}
&\frac{\partial Q_3(\boldsymbol{\vartheta}_Y\mid \boldsymbol{\theta}^{(m-1)})}{\partial \boldsymbol{\Gamma}^k_{*}}=\mathbf{0},\quad \frac{\partial Q_3(\boldsymbol{\vartheta}_Y\mid \boldsymbol{\theta}^{(m-1)})}{\partial \boldsymbol{\alpha}_{Y,k}}=\mathbf{0},\\
&-\sum_{i=1}^n t_{ik}^{(m)}wi_{ik,Y}^{(m)}\boldsymbol{\Sigma}_{Y,k}^{-1}\left(-\mathbfit{c}_{Y,i}({\mathbfit{c}_{X,i}^{*}})^\top+\boldsymbol{\Gamma}^k_{*}{\mathbfit{c}_{X,i}^{*}}({\mathbfit{c}_{X,i}^{*}})^\top\right)-\sum_{i=1}^n t_{ik}^{(m)}\boldsymbol{\Sigma}_{Y,k}^{-1}\boldsymbol{\alpha}_{Y,k}{\mathbfit{c}_{X,i}^{*}}^\top=\mathbf{0}.\\
&-\sum_{i=1}^n t_{ik}^{(m)}w_{ik,Y}^{(m)}\boldsymbol{\Sigma}_{Y,k}^{-1}\boldsymbol{\alpha}_{Y,k}+\sum_{i=1}^n t_{ik}\biggl(\boldsymbol{\Sigma}_{Y,k}^{-1}\mathbfit{c}_{Y,i}-\boldsymbol{\Sigma}_{Y,k}^{-1}\boldsymbol{\Gamma}^k_{*}{\mathbfit{c}_{X,i}^{*}}\biggl)=\mathbf{0}.
\end{align*}

Notice that using again properties of trace and transpose we have
\begin{align*}
&Q_3(\boldsymbol{\vartheta}_Y\mid \boldsymbol{\theta}^{(m-1)})=-\frac{1}{2}\sum_{i=1}^n\sum_{k=1}^K t_{ik}^{(m)}lw_{ik,Y}^{(m)}-\frac{nR_Y\log(2\pi)}{2}-\frac{1}{2}\sum_{k=1}^K n_{k}^{(m)}\log(\mid \boldsymbol{\Sigma}_{Y,k}\mid)  \\
&-\frac{1}{2}\sum_{i=1}^n\sum_{k=1}^K t_{ik}^{(m)}wi_{ik,Y}^{(m)}{\text{trace}}\left(({\mathbfit{c}}_{Y,i}-\boldsymbol{\Gamma}^k_{*}{\mathbfit{c}_{X,i}^{*}} )^\top\boldsymbol{\Sigma}_{Y,k}^{-1}({\mathbfit{c}}_{Y,i}-\boldsymbol{\Gamma}^k_{*}{\mathbfit{c}_{X,i}^{*}} )\right)\\
&-\frac{1}{2}\sum_{i=1}^n\sum_{k=1}^K t_{ik}^{(m)}w_{ik,Y}^{(m)}{\text{trace}}\left(\boldsymbol{\alpha}_{Y,k}^\top\boldsymbol{\Sigma}_{Y,k}^{-1}\boldsymbol{\alpha}_{Y,k}\right)\\
&+\frac{1}{2}\sum_{i=1}^n\sum_{k=1}^K t_{ik}^{(m)}\left({\text{trace}}\left(\left(\mathbfit{c}_{Y,i}
-\boldsymbol{\Gamma}^k_{*}{\mathbfit{c}_{X,i}^{*}}\right)^\top\boldsymbol{\Sigma}_{Y,k}^{-1}\boldsymbol{\alpha}_{Y,k}\right)
+{\text{trace}}\left(\boldsymbol{\alpha}_{Y,k}^\top\boldsymbol{\Sigma}_{Y,k}^{-1}\left(\mathbfit{c}_{Y,i}-\boldsymbol{\Gamma}^k_{*}{\mathbfit{c}_{X,i}^{*}}\right)\right)\right)\\
&=-\frac{1}{2}\sum_{i=1}^n\sum_{k=1}^K t_{ik}^{(m)}lw_{ik,Y}^{(m)}-\frac{nR_Y\log(2\pi)}{2}-\frac{1}{2}\sum_{k=1}^K n_{k}^{(m)}\log(\mid \boldsymbol{\Sigma}_{Y,k}\mid)\\
 &-\frac{1}{2}\sum_{i=1}^n\sum_{k=1}^K t_{ik}^{(m)}wi_{ik,Y}^{(m)}{\text{trace}}\left(\boldsymbol{\Sigma}_{Y,k}^{-1}({\mathbfit{c}}_{Y,i}-\boldsymbol{\Gamma}^k_{*}{\mathbfit{c}_{X,i}^{*}} )({\mathbfit{c}}_{Y,i}-\boldsymbol{\Gamma}^k_{*}{\mathbfit{c}_{X,i}^{*}} )^\top\right)\\
 &-\frac{1}{2}\sum_{i=1}^n\sum_{k=1}^K t_{ik}^{(m)}w_{ik,Y}^{(m)}{\text{trace}}\left(\boldsymbol{\Sigma}_{Y,k}^{-1}\boldsymbol{\alpha}_{Y,k}\boldsymbol{\alpha}_{Y,k}^\top\right)\\
 &+\frac{1}{2}\sum_{i=1}^n\sum_{k=1}^K t_{ik}^{(m)}\left({\text{trace}}\left(\boldsymbol{\Sigma}_{Y,k}^{-1}\left(\mathbfit{c}_{Y,i}
-\boldsymbol{\Gamma}^k_{*}{\mathbfit{c}_{X,i}^{*}}\right)\boldsymbol{\alpha}_{Y,k}^\top\right)
+{\text{trace}}\left(\boldsymbol{\Sigma}_{Y,k}^{-1}\boldsymbol{\alpha}_{Y,k}\left(\mathbfit{c}_{Y,i}-\boldsymbol{\Gamma}^k_{*}{\mathbfit{c}_{X,i}^{*}}\right)\right)^\top\right)
\end{align*}
We obtain formula \eqref{sigmY} solving 
\begin{align*}
&\frac{\partial Q_3(\boldsymbol{\vartheta}_Y\mid \boldsymbol{\theta}^{(m-1)})}{\partial \boldsymbol{\Sigma}_{Y,k}^{-1}}=\mathbf{0}.
\end{align*}

If $\mathbfit{c}_{X,i}\mid Z_{ik}=1\sim VG_{R^X}(\boldsymbol{\mu}_{X,k},\boldsymbol{\alpha}_{X,k},\boldsymbol{\Sigma}_{X,k},\psi_{X,k})$, $k=1,\ldots, K$, then
\begin{align*}
&Q_{41}(\psi_{X,1},\ldots, \psi_{X,K}\mid\boldsymbol{\theta}^{(m-1)}):=\sum_{i=1}^n\sum_{k=1}^KE[ z_{ik}\log (h(w_{i,X};\theta_{W,X}))\mid \mathbfit{c}_{X,1},\ldots,\mathbfit{c}_{X,n},\boldsymbol{\theta}^{(m-1)} ]\\
&=\sum_{i=1}^n\sum_{k=1}^Kt_{ik}^{(m)}\biggl( \psi_{X,k}\log(\psi_{X,k})-\log({\Gamma(\psi_{X,k})}+(\psi_{X,k}-1)lw_{ik,X}^{(m)}-\psi_{X,k}w_{ik,X}^{(m)}\biggl)
\end{align*}
Solving 
$$
\frac{\partial Q_{41}(\psi_{X,1},\ldots, \psi_{X,K}\mid\boldsymbol{\theta}^{(m-1)})}{\partial\psi_{X,k}}=0, \quad k=1,\ldots, K,
$$
we obtain that the update $\psi_{X,k}^{(m)}$ is the solution of the equation \eqref{specparVGx1}.

If $\mathbfit{c}_{X,i}\mid Z_{ik}=1\sim ST_{R^X}(\boldsymbol{\mu}_{X,k},\boldsymbol{\alpha}_{X,k},\boldsymbol{\Sigma}_{X,k},\nu_{X,k})$, $k=1,\ldots, K$, then
\begin{align*}
&Q_{41}(\psi_{X,1},\ldots, \psi_{X,K}\mid\boldsymbol{\theta}^{(m-1)}):=\sum_{i=1}^n\sum_{k=1}^KE[ z_{ik}\log (h(w_{i,X};\theta_{W,X}))\mid \mathbfit{c}_{X,1},\ldots,\mathbfit{c}_{X,n},\boldsymbol{\theta}^{(m-1)} ]\\
&=\sum_{i=1}^n\sum_{k=1}^Kt_{ik}^{(m)}\biggl( \frac{\nu_{X,k}}{2}\log\left(\frac{\nu_{X,k}}{2}\right)-\log\left({\Gamma\left(\frac{\nu_{X,k}}{2}\right)}\right)-\left(\frac{\nu_{X,k}}{2}+1\right)lw_{ik,X}^{(m)}-\frac{\nu_{X,k}}{2}wi_{ik,X}^{(m)}\biggl)
\end{align*}
Solving 
$$
\frac{\partial Q_{41}(\psi_{X,1},\ldots, \psi_{X,K}\mid\boldsymbol{\theta}^{(m-1)})}{\partial\nu_{X,k}}=0, \quad k=1,\ldots, K,
$$
we obtain that the update $\nu_{X,k}^{(m)}$ is the solution of the equation \eqref{specparSTx1}.

If $\mathbfit{c}_{X,i}\mid Z_{ik}=1\sim NIG_{R^X}(\boldsymbol{\mu}_{X,k},\boldsymbol{\alpha}_{X,k},\boldsymbol{\Sigma}_{X,k},\kappa_{X,k})$, $k=1,\ldots, K$, then
\begin{align*}
&Q_{41}(\psi_{X,1},\ldots, \psi_{X,K}\mid\boldsymbol{\theta}^{(m-1)}):=\sum_{i=1}^n\sum_{k=1}^KE[ z_{ik}\log (h(w_{i,X};\theta_{W,X}))\mid \mathbfit{c}_{X,1},\ldots,\mathbfit{c}_{X,n},\boldsymbol{\theta}^{(m-1)} ]\\
&=\sum_{i=1}^n\sum_{k=1}^Kt_{ik}^{(m)}\biggl( -\frac{1}{2}\log(2\pi)+\kappa_{X,k}-\frac{3}{2}lw_{ik,X}^{(m)}-\frac{1}{2}wi_{ik,X}^{(m)}-\frac{1}{2}\kappa_{X,k}^2w_{ik,Y}^{(m)}\biggl)
\end{align*}
Solving 
$$
\frac{\partial Q_{41}(\psi_{X,1},\ldots, \psi_{X,K}\mid\boldsymbol{\theta}^{(m-1)})}{\partial\kappa_{X,k}}=0, \quad k=1,\ldots, K,
$$
we obtain \eqref{specparNIGx1}.
\end{proof}
\section{Coefficients of the simulated data}
\label{apendD}
For the NIG-NIG, ST-ST and NIG-VG simualtions we put 
\begin{align*}
&\boldsymbol{\Sigma}_{X,1}=\begin{pmatrix}
35293.603& -34652.17& 33635.415& -21127.07&  16185.294&  6203.499\\
-34652.172&  53300.21& -50164.778&  39232.83& -23938.415& 11415.853\\
33635.415& -50164.78&  57345.569& -45236.01&  32712.598& -5203.224\\
-21127.071&  39232.83& -45236.014&  42343.90& -30233.209& 16558.252\\
16185.294& -23938.42&  32712.598& -30233.21&  28223.659& -7432.215\\
6203.499&  11415.85&  -5203.224&  16558.25&  -7432.215& 37585.733
\end{pmatrix},\\
&\boldsymbol{\Sigma}_{X,2}=\begin{pmatrix}
32496.743& -33954.84&  30255.33& -19930.28&  14794.720&   2833.563\\
-33954.843&  53968.88& -50455.29&  40307.52& -23295.369&  14702.169\\
30255.329& -50455.29&  54784.92& -45636.53&  30633.456& -12566.028\\
-19930.283&  40307.52& -45636.53&  44069.51& -29155.230&  21986.692\\
14794.720& -23295.37&  30633.46& -29155.23&  27636.508&  -8247.361\\
2833.563&  14702.17& -12566.03&  21986.69 & -8247.361&  40627.008
\end{pmatrix},\\
&\boldsymbol{\Gamma}^1=\begin{pmatrix}
-0.2425716&  0.54897465& -0.8916372&  1.29210175& -2.1477465& 3.3094321\\
-0.1060100&  1.26481276& -1.9308257&  2.36947593& -3.4423522& 4.2985155\\
-1.1520068&  0.93679861& -2.2869660&  3.93104866& -5.8769335& 7.0047824\\
4.3109023&  0.09308357& -1.7355627&  0.16265468&  0.3923705& 0.4745340\\
0.6957114&  0.80979318& -1.6477535&  1.49785833& -1.3492138& 1.6052517\\
3.3165545& -1.96173600&  0.4144113& -0.05521575& -0.1246685& 0.4763175
\end{pmatrix},
\end{align*}
\begin{align*}
&\boldsymbol{\Gamma}^2=\begin{pmatrix}
-0.18286204&  0.4846445& -0.8103231&  1.1106091 &-1.8322473& 3.0437840\\
-0.09704194&  0.9753574& -1.5860429&  1.9259053 &-2.8032265& 3.8551773\\
-1.21799403&  0.8923668& -1.4619894&  2.5499362& -4.7798136& 6.5881388\\
1.45078879&  2.6988567& -4.0145100&  2.2191593& -0.5766956& 0.8673926\\
-0.29982902&  0.9611298& -0.7879001&  0.4851317& -0.3288526& 0.8183521\\
2.00917573& -1.0346535&  0.5647058& -0.5797621&  0.4183034& 0.1227972
\end{pmatrix},
\end{align*}
$\boldsymbol{\Gamma}_0^1=(4.978059,-150.127321, 294.147021, -803.866942, 57.684388, 211.959684)^\top$, \\
$\boldsymbol{\Gamma}_0^2=(0.1084669,-59.2740539, 302.7418344,-1012.3411351, 111.9925126, 172.9547505)^\top$.\\
For the NIG-VG scenario we have:\\
 $\boldsymbol{\mu}_{X,1}=(763.1701, 679.3222, 465.8823, 544.5796, 640.5101, 642.5667)^\top$, \\
  $\boldsymbol{\mu}_{X,2}=(778.8822, 750.8995, 402.3499, 836.9349, 840.3188, 831.0520)^\top$.\\
 For the NIG-NIG and ST-ST simulations we use\\
  $\boldsymbol{\mu}_{X,1}=(1526.340, 1358.644, 931.7646, 1089.159, 1281.020, 1285.133)^\top$, \\
   $\boldsymbol{\mu}_{X,2}=(1557.764, 1501.799, 804.6998, 1673.870, 1680.638, 1662.104)6\top$. 
   
  For the VG-VG   simulations we have
   \begin{align*}
&\boldsymbol{\Sigma}_{Y,1}=\boldsymbol{\Sigma}_{Y,2}=\begin{pmatrix}
28.35493&  28.62064  &118.3307&   95.89802 & 42.41898 & 36.26409\\
28.62064 &226.48897&  150.7904&  371.04226& 186.19665& 134.65827\\
118.33066 &150.79045& 1241.6319&  549.88239& 412.62674& 259.10875\\
 95.89802& 371.04226&  549.8824 &2616.75870& 836.74973& 835.79828\\
42.41898 &186.19665&  412.6267 & 836.74973& 749.32405& 404.91274\\
36.26409 &134.65827  &259.1088&  835.79828 &404.91274 &412.15975
\end{pmatrix},
\end{align*}
 \begin{align*}
&\boldsymbol{\Sigma}_{X,1}=\begin{pmatrix}
35293.603& -34652.17&  33635.415& -21127.07&  16185.294&  6203.499\\
-34652.172&  53300.21 &-50164.778&  39232.83 &-23938.415& 11415.853\\
33635.415& -50164.78&  57345.569& -45236.01&  32712.598 &-5203.224\\
-21127.071 & 39232.83& -45236.014&  42343.90 &-30233.209& 16558.252\\
16185.294 &-23938.42&  32712.598& -30233.21 & 28223.659& -7432.215\\
6203.499&  11415.85 & -5203.224 & 16558.25&  -7432.215& 37585.733
\end{pmatrix},\\
&\boldsymbol{\Sigma}_{X,2}=\begin{pmatrix}
32496.743& -33954.84&  30255.33& -19930.28 & 14794.720&   2833.563\\
-33954.843 & 53968.88& -50455.29&  40307.52& -23295.369  &14702.169\\
30255.329& -50455.29&  54784.92& -45636.53&  30633.456& -12566.028\\
-19930.283&  40307.52& -45636.53&  44069.51& -29155.230&  21986.692\\
14794.720& -23295.37&  30633.46& -29155.23&  27636.508&  -8247.361\\
 2833.563 & 14702.17 &-12566.03&  21986.69 & -8247.361&  40627.008
\end{pmatrix},\\
&\boldsymbol{\Gamma}^1=\begin{pmatrix}
-0.2425716&  0.54897465& -0.8916372&  1.29210175& -2.1477465 &3.3094321\\
-0.1060100&  1.26481276 &-1.9308257&  2.36947593& -3.4423522 &4.2985155\\
-1.1520068&  0.93679861& -2.2869660 & 3.93104866& -5.8769335 &7.0047824\\
4.3109023&  0.09308357 &-1.7355627&  0.16265468&  0.3923705& 0.4745340\\
0.6957114&  0.80979318& -1.6477535&  1.49785833& -1.3492138 &1.6052517\\
3.3165545& -1.96173600&  0.4144113& -0.05521575& -0.1246685& 0.4763175
\end{pmatrix},\\
&\boldsymbol{\Gamma}^2=\begin{pmatrix}
-0.18286204&  0.4846445& -0.8103231&  1.1106091& -1.8322473& 3.0437840\\
-0.09704194&  0.9753574 &-1.5860429&  1.9259053& -2.8032265 &3.8551773\\
-1.21799403&  0.8923668& -1.4619894 & 2.5499362 &-4.7798136 &6.5881388\\
1.45078879&  2.6988567& -4.0145100&  2.2191593 &-0.5766956 &0.8673926\\
 -0.29982902&  0.9611298 &-0.7879001&  0.4851317 &-0.3288526 &0.8183521\\
2.00917573& -1.0346535&  0.5647058 &-0.5797621&  0.4183034& 0.1227972
\end{pmatrix},
\end{align*}
$\boldsymbol{\Gamma}_0^1=(4.978059, -150.127321, 294.147021, -803.866942, 57.684388, 211.959684)^\top$, \\
$\boldsymbol{\Gamma}_0^2=( 0.1084669, -59.2740539, 302.7418344, -1012.3411351,111.9925126, 172.9547505)^\top$.\\
$\boldsymbol{\mu}_{X,1}=(763.1701, 679.3222, 465.8823, 544.5796, 640.5101, 642.5667)^\top$, \\
  $\boldsymbol{\mu}_{X,2}=( 2 778.8822 750.8995 402.3499 836.9349 840.3188 831.0520)^\top$.\\
\subsection{Mean Square Errors}
The mean square errors for the entries in the $\boldsymbol{\Gamma}^1$, $\boldsymbol{\Gamma}^2$ matrices for the thresholds $\epsilon$ that give the largest ARI  are
\begin{itemize}
\item[NIG-VG] 
\begin{align*}
&\boldsymbol{\Gamma}^1=\begin{pmatrix}
0.05511158& 0.06750058& 0.03966146& 0.05387034& 0.12635590& 0.08697222\\
0.05625283& 0.08496746& 0.05573041& 0.07208543& 0.08821545&0.08022848\\
0.07203235 &0.07370440& 0.10107496 &0.16392964& 0.15243120 &0.16731654\\
0.18335635& 0.25164329& 0.18566190& 0.16862247& 0.17801753& 0.21055817\\
0.12450012& 0.06226461& 0.09511584& 0.11862939 &0.12978367& 0.09830321\\
0.09357647& 0.13245603& 0.05517701& 0.10658585& 0.17038248& 0.15353957
\end{pmatrix},\\
&\boldsymbol{\Gamma}^2=\begin{pmatrix}
0.04676092& 0.51061654& 0.5629539& 0.5218262& 0.5406844 &0.29309801\\
0.06049283& 0.21977154 &0.2694277 &0.1516302 &0.1462645 &0.20768454\\
0.05515323 &0.49313048& 0.5962365& 0.4619501& 0.4821558 &0.38064281\\
0.35075613& 0.59334703& 0.8853335& 0.9324366 &1.0983315& 0.73714940\\ 
0.11899463& 0.63223866 &0.8198979& 0.6945194& 0.7651928& 0.55392985\\
0.19884166& 0.06777988 &0.1448004& 0.1317819& 0.1093196& 0.08511835
\end{pmatrix},
\end{align*} 
\item[NIG-NIG] 
\begin{align*}
&\boldsymbol{\Gamma}^1=\begin{pmatrix}
0.04949456& 0.05267041 &0.05171151& 0.06433869& 0.06642077& 0.04645552\\
0.05344229& 0.04795297 &0.05051225& 0.05924572& 0.06339355 &0.04111531\\
0.05298855& 0.06259288& 0.05849245& 0.08903793& 0.07209674 &0.05453267\\
 0.08438712 &0.15459287& 0.10507622& 0.17186504& 0.15605905& 0.09311142\\
0.04872996 &0.07227091& 0.06382920& 0.08980058& 0.08037638 &0.04788885\\
0.05646893& 0.08387114& 0.06353995& 0.09899717 &0.08674788& 0.05551675
\end{pmatrix},\\
&\boldsymbol{\Gamma}^2=\begin{pmatrix}
0.05051562 &0.04782073& 0.05190849& 0.05949210& 0.06115799& 0.05115529\\
0.05083599 &0.04964070& 0.04885193& 0.05624939 &0.06299182& 0.05579979\\
0.05024124 &0.05223636 &0.05173883& 0.07246345 &0.06544097 &0.05227624\\
0.08447716& 0.11065377& 0.07232235& 0.10045061& 0.11040111& 0.08256365\\
0.05168826& 0.06060290 &0.05579542& 0.07186421& 0.06678256& 0.05461314\\
0.05804121& 0.07868333 &0.06731626& 0.08382073& 0.08103683 &0.06377735
\end{pmatrix},
\end{align*} 
\item[ST-ST] 
\begin{align*}
&\boldsymbol{\Gamma}^1=\begin{pmatrix}
0.03390765 &0.03163137& 0.03922696& 0.03828476& 0.04176651& 0.03878768\\
0.04320439 &0.03871281 &0.03951894& 0.05816802& 0.05885902 &0.06050412\\
0.03598567& 0.03350200 &0.03324511& 0.05396178& 0.03970275& 0.04475850\\
0.05087718& 0.04521354& 0.04330248& 0.08625664& 0.06851708& 0.08540129\\
0.04273736& 0.03592602 &0.03884678& 0.05503644 &0.04335571& 0.04977443\\
0.03555987& 0.03833964& 0.04194613 &0.04860137& 0.05344222& 0.05036222
\end{pmatrix},\\
&\boldsymbol{\Gamma}^2=\begin{pmatrix}
0.03424853& 0.03274469& 0.04422438& 0.04466440 &0.04825152 &0.04690288\\
0.03627615& 0.03533435& 0.03713555& 0.04427395 &0.05074950& 0.04403081\\
0.03679642& 0.03688324& 0.03639976& 0.05034130& 0.04796767& 0.04659711\\
0.04045040& 0.04337199& 0.04248414& 0.07708740& 0.05477124& 0.07035285\\
0.04014917& 0.03781302 &0.04090921& 0.06359574& 0.05419604 &0.06241279\\
0.03570126& 0.03305470& 0.03547626& 0.05594767& 0.05261839& 0.05079321
\end{pmatrix},
\end{align*} 
\item[VG-VG] 
\begin{align*}
&\boldsymbol{\Gamma}^1=\begin{pmatrix}
0.01755966 &0.01950570& 0.02710265& 0.03348576& 0.03373675& 0.03074773\\
0.02829054& 0.05194753& 0.03933944 &0.04401189& 0.06268247 &0.07360047\\
0.07244793& 0.08317489& 0.07199904& 0.09910160& 0.11902539& 0.14622145\\
0.08053016 &0.09915225& 0.12044524& 0.17313319 &0.17274423& 0.22254503\\
0.04417642& 0.03698360 &0.04971864& 0.06055095& 0.05317780& 0.06029568\\
0.04721025& 0.05356217 &0.06708498 &0.06382525& 0.07873284& 0.09756050
\end{pmatrix},\\
&\boldsymbol{\Gamma}^2=\begin{pmatrix}
0.009431315 &0.008066421& 0.007942953& 0.01001032 &0.01295720& 0.01447825\\
0.026586007& 0.025338466 &0.024038376 &0.03377267& 0.03078340& 0.03712832\\
0.057705259 &0.056472504& 0.059208995& 0.07506966& 0.07482930& 0.07631925\\
0.149216996& 0.142586981& 0.112561286& 0.16697032 &0.11927667& 0.21081798\\
0.069983266& 0.042186203& 0.045541297& 0.06717144& 0.05969463& 0.06481972\\
0.068802443& 0.049478458& 0.039891521& 0.07188424& 0.05133695& 0.07072622
\end{pmatrix},
\end{align*} 
\end{itemize}
\end{appendices}



\end{document}